\documentclass[paper]{JHEP3}
\usepackage{amssymb,amsmath}
\input{xy}
\xyoption{all}

\DeclareMathOperator{\sign}{\text{sign}}
\DeclareMathOperator{\End}{\text{End}}
\DeclareMathOperator{\Inv}{\text{Inv}}

\newcommand{\bartial}{{\bar{\partial}}}
\newcommand{\Real}{\mathbb{R}}

\newcommand{\Integer}{\mathbb{Z}}

\newcommand{\mat}{\begin{pmatrix}}
\newcommand{\tam}{\end{pmatrix}}
\newcommand{\smat}{\left(\begin{smallmatrix}}
\newcommand{\stam}{\end{smallmatrix}\right)}
\newcommand{\mc}{\mathcal}

\newcommand{\affH}{{\hat{H}_4}}

\newcommand{\HHH}{{\hat{H}_4\times\hat{H}_4/\hat{H}_4}}

\newcommand{\nubb}{\bar{\nu}}
\newcommand{\mubb}{{\bar{\mu}}}
\newcommand{\etabb}{{\bar{\eta}}}
\newcommand{\zetab}{{\bar{\zeta}}}
\newcommand{\balpha}{\bar{\alpha}}

\newcommand{\ad}{{\text{ad}}}
\newcommand{\tr}{\text{tr}}

\newcommand{\g}{\mathfrak{g}}
\newcommand{\h}{\mathfrak{h}}

\newcommand{\sym}{{\text{sym}}}

\newcommand{\be}{\begin{equation}}
\newcommand{\ee}{\end{equation}}
\newcommand{\ba}{\begin{eqnarray}}
\newcommand{\ea}{\end{eqnarray}}
\newcommand{\F}{\Phi}

\newcommand{\e}{\epsilon}

\newcommand{\m}{\mu}
\newcommand{\n}{\nu}
\newcommand{\s}{\sigma}

\newcommand{\p}{\partial}

\newcommand{\f}{\varphi}

\newcommand{\w}{\omega}

\newcommand{\nb}{\nonumber}

\title{The diagonal cosets of the Heisenberg group}

\author{Giuseppe D'Appollonio${}^{a,b}$ and Thomas Quella${}^c$ \\
  ${}^a$ Department of Mathematics, King's College London\\
The Strand, London WC2R 2LS, United Kingdom \vspace{0.2cm} \\
${}^b$ Dipartimento di Fisica dell'Universit\`a di Cagliari, INFN Sezione di Cagliari\\
Cittadella Universitaria 09042 Monserrato, Italy  \vspace{0.2cm} \\
${}^c$ Institute for Theoretical Physics, University of
Amsterdam\\ Valckenierstraat 65, 1018 XE Amsterdam, The Netherlands  }

\preprint{NI08011-SIS \\ ITFA-2007-57 \\ KCL-MTH-07-17 \\ arXiv:0801.4634}

\abstract{
   In this paper we study the diagonal cosets of the non-compact
  $H_4$ WZW model. Generalising earlier work by
  Antoniadis and Obers, we provide an exact world-sheet description
  for several families of non-maximally symmetric gravitational plane
  waves with background NS fluxes. We show that the $\sigma$-models
  that correspond to an asymmetric action of the gauge group smoothly interpolate
  between singular and non-singular plane waves. We also analyse the
  representations of the coset chiral algebra and derive the
  spectrum of all the models.
}

\keywords{Conformal Field Models in String Theory, Conformal and W Symmetry, Bosonic Strings}

\begin{document}

\section{Introduction}

Curved string backgrounds provide a well-defined context to explore
the properties of string theory as a theory of quantum gravity. In
certain cases string propagation in these spaces can even provide an
holographic description of the dynamics of a dual gauge theory
\cite{Aharony:1999ti}. For these reasons one would like to enlarge the
class of curved space-times for which an exact conformal field theory
description is available as much as possible. With this aim in mind,
we started in \cite{D'Appollonio:2007bn} a systematic study of the
non-compact coset models based on the Heisenberg group
$H_4$. Non-compact cosets are a very interesting class of string
theory backgrounds since they can be studied from two complementary
points of view. On the one hand their Lagrangian formulation as gauged WZW
models
\cite{Bardakci:1987ee,Gawedzki:1988hq,Gawedzki:1989nj,Karabali:1989au,Karabali:1990dk,Bardakci:1990wj}
provides a clear space-time interpretation, on the other hand their
exact conformal field theory description \cite{Bardakci:1971nb,
  Goddard:1985vk} allows to derive the spectrum of string excitations
and calculate their scattering amplitudes.

  Using the geometric formulation, several curved backgrounds were
  soon recognised as non-compact coset theories: the two-dimensional
  black-hole \cite{Witten:1991yr}, the three-dimensional black string
  \cite{Horne:1991gn}, the inhomogeneous Nappi-Witten cosmology
  \cite{Nappi:1992kv} as well as many other examples
  \cite{Kounnas:1992wc, Ginsparg:1992af}. On the other hand for
  several years it was not possible to use conformal field theory
  techniques to derive the spectrum and compute the interactions,
  since the representation theory of the non-compact affine algebras
  was not properly understood and their structure constants were not
  known. The situation changed with the work of Teschner as well as
  Maldacena and Ooguri \cite{Teschner:1997ft, Maldacena:2000hw,
    Maldacena:2001km}. These authors clarified the operator content of
  the $SL(2, \mathbb{R})$ WZW model and of its Euclidean analogue, the
  $H_3^+$ model, and derived their structure constants. Using these
  results, it was finally possible to analyse in some detail the
  conformal field theories of abelian cosets based on
  $SL(2,\mathbb{R})$ \cite{Maldacena:2000kv,Elitzur:2002rt,
    Giveon:2003ge,Ribault:2003ss}.

  As is well-known, the $SL(2,\mathbb{R})$ WZW model describes the
  propagation of strings in $AdS_3$. The Heisenberg group $H_4$
  considered in the present paper, a contraction of
  $SL(2,\mathbb{R})\times U(1)$, describes the propagation of strings
  in a four-dimensional, maximally symmetric plane wave with seven
  isometries \cite{Nappi:1993ie}. The study of the representation
  theory of the $H_4$ affine algebra was started in
  \cite{Kiritsis:1993jk} and the model was exactly solved in
  \cite{D'Appollonio:2003dr, Bianchi:2004vf}. Also the exact solution
  of the boundary CFTs of the maximally symmetric D-branes in this
  background is available \cite{D'Appollonio:2004pm}.

  In our first paper \cite{D'Appollonio:2007bn} we considered the
  abelian cosets of the Heisenberg group \cite{Kiritsis:1993jk,
    Sfetsos:1993rh, Sfetsos:1994fc} and showed that they provide a CFT
  description of several three-dimensional backgrounds such as the
  Melvin model \cite{Tseytlin:1994ei}, the conical point-particle
  space-times \cite{Deser:1983tn} and the null orbifold
  \cite{Liu:2002ft}. In the present paper we perform a detailed study
  of the diagonal cosets of the Heisenberg group which exhibit a
  number of interesting new features. For these examples both the
  numerator and the denominator group of the coset are non-compact and
  non-abelian. To our knowledge such theories have hardly been studied
  from an algebraic point of view so far.

  As shown by Antoniadis and Obers these models have also a very
  interesting geometric interpretation in terms of non-maximally
  symmetric plane waves \cite{Antoniadis:1994jx}. In the paper
  mentioned above the authors discussed two special classes of
  diagonal cosets. The first class is given by a two-parameter family
  of singular geometries which are T-dual to plane gravitational
  waves. The second class describes a one-parameter family of
  gravitational waves with five isometries. Although it is well-known
  that $\s$-models associated with plane gravitational waves are
  always conformally invariant if the dilaton and the three-form flux
  are chosen appropriately
  \cite{Gueven:1987ad,Amati:1988ww,Horowitz:1989bv,Horowitz:1990sr,Tseytlin:1992pq},
  the identification of the underlying conformal field theories is a
  highly non-trivial task. The work of Antoniadis and Obers provided
  this identification for a whole class of plane wave backgrounds.

  In the present paper we generalise the analysis of Antoniadis and
  Obers in two respects. Firstly we show that by relaxing a few
  assumptions the diagonal cosets encompass further string backgrounds
  that have not been considered so far. In addition we go beyond the
  pure Lagrangian and geometric description and make a significant
  step towards a full conformal field theory analysis of these
  models. One of the main results of our paper is the classification
  of all possible diagonal cosets of the Heisenberg group and the
  explicit construction of the corresponding $\s$-models. Since $H_4$
  is non-semisimple, it admits continuous families of non-isometric
  automorphisms, and each of them may be used to deform the standard
  diagonal embedding. Combined with the freedom of choosing different
  embeddings of $H_4$ in the left and the right sectors of the
  original CFT the existence of these continuous families of
  automorphisms leads to an extremely rich number of possible coset
  models. They can be divided into three classes that we will refer to
  as $(++)$, $(+-)$ and $(--)$. Each class depends on several
  continuous parameters. For certain restricted choices of the
  parameters of the $(++)$ and $(--)$ classes, one recovers the models
  constructed by Antoniadis and Obers.

  In this paper we also present the derivation of the spectrum of the
  diagonal cosets using conformal field theory techniques. In order to
  achieve this goal we study the decomposition of the tensor products of
  affine $H_4$ representations with respect to the embedded $H_4$
  algebra. In contrast to compact or abelian cosets the standard
  method of determining the branching functions fails since products
  of affine characters are usually divergent if considered as
  generating functions for the states of a representation.
  This problem already shows up at the level of the horizontal
  subalgebra, since the non-trivial unitary representations of a
  non-compact group are infinite dimensional.

  To circumvent this problem we develop a new method for the derivation of the affine coset
  characters which makes use both of character decompositions and of the
  knowledge of the tensor products of the horizontal subalgebra. The
  latter should be thought of as providing some analytical input which
  allows to deal with the mathematical difficulties of having infinite
  dimensional weight spaces. While we will be able to provide a
  complete answer for the decomposition of the tensor products of
  highest weight representations, we only have partial results for the
  tensor products of spectral flow representations. The fact that the
  adjoint representation of $H_4$ is indecomposable but reducible
  potentially leads to further complications.

  This paper is organised as follows. In section~\ref{sc:Data} we
  begin with a brief review of the construction of asymmetrically
  gauged WZW models. After a classification of the diagonal embeddings
  of the Lie algebra $H_4$ we determine all possible diagonal
  cosets and derive the quantities needed for their Lagrangian
  description. In section~\ref{sc:CosetGeometries} we present the
  metric, the dilaton and the antisymmetric tensor for our three
  classes of models. The background data are displayed only for two
  particular families of parameters while the most general expressions
  are collected in appendix~\ref{ap:FullGeometries}. In
  section~\ref{sc:RepTh} we proceed to a more algebraic treatment and
  derive explicit formulas for the diagonal coset characters. These
  results are used in section~\ref{sc:Spectrum} to compute the
  spectrum of the three classes of diagonal
  cosets. Section~\ref{sc:conclusions} contains our conclusions and
  some comments on possible extensions of our work.

\section{\label{sc:Data}Classification of the diagonal cosets}

  After a brief review of the Lagrangian description of the gauged WZW
  models, we classify all possible diagonal cosets of the
  Heisenberg group. We thereby generalise the analysis of
  \cite{Antoniadis:1994jx} and provide the grounds for a thorough
  treatment of the coset geometries to be described in section
  \ref{sc:CosetGeometries}.

\subsection{Asymmetrically gauged Wess-Zumino-Witten models}

  Let $G$ be a Lie group and $H$ a Lie subgroup of $G$. A general
  coset model $G/H$ is completely specified by the choice of two
  invariant forms $\langle\cdot,\cdot\rangle_G$ and
  $\langle\cdot,\cdot\rangle_H$ on the respective Lie algebras and the
  selection of two embeddings $\epsilon, \bar \e:H\to G$. The coset
  space is then determined by the identification
\begin{equation}
  G/H\ =\ \bigl\{g\in G\bigl|\,g\sim\epsilon(h)g \bar \epsilon(h^{-1}),
  \forall \, h \, \in H \,\bigl\}\ \ .
\end{equation}
  Provided that the consistency
  requirement
\begin{equation}
  \label{eq:NoAnomaly}
   \bigl\langle\epsilon (X),\epsilon (Y)\bigr\rangle_G
  \ =\ \bigl\langle \bar \epsilon (X),\bar \epsilon (Y)\bigr\rangle_G
  \ =\ \bigl\langle X,Y\bigr\rangle_H\quad\text{ for all }X,Y\in\h \
   ,
\end{equation}
  is satisfied, these data define a conformally invariant
  $\sigma$-model on $G/H$ via the construction of gauged Wess-Zumino-Witten models
  \cite{Bardakci:1987ee,Gawedzki:1988hq,Gawedzki:1989nj,Karabali:1989au,Karabali:1990dk,Bardakci:1990wj}.

  The starting point of this construction is the action
\begin{equation}
  \label{eq:GeneralCoset}
  \begin{split}
    \mc{S}^{G/H}(g,U,V)\ =\ \mc{S}^G\bigl(\epsilon(U^{-1}) \, g \, \bar \epsilon(V)\bigr)-\mc{S}^H(U^{-1}V)\ \ ,
  \end{split}
\end{equation}
  where $g:\Sigma\to G$ and $U,V:\Sigma\to H$ are group valued
  fields. Here the symbol $\mc{S}^G$ denotes the WZW Lagrangian for
  the group $G$
\begin{equation}
  \label{eq:GeneralWZW}
  \mc{S}^G(g)
  \ =\ -\frac{i}{4\pi}\int\bigl\langle g^{-1}\partial g,g^{-1}\bartial g\bigr\rangle_G\,dz\wedge d\bar{z}
       -\frac{i}{24\pi}\int\bigl\langle
       g^{-1}dg,[g^{-1}dg,g^{-1}dg]\bigr\rangle_G \ ,
\end{equation}
  and similarly $\mc{S}^H$ denotes the WZW Lagrangian for the group
  $H$. The action \eqref{eq:GeneralCoset} is manifestly invariant
  under local $H$-transformations of the form
\begin{align}
  g&\mapsto \epsilon(h)g\bar \epsilon(h^{-1})\ ,&
  U&\mapsto hU\ ,&
  V&\mapsto hV\ \ .
\end{align}
  To further simplify the action \eqref{eq:GeneralCoset}, we introduce
  the gauge fields $\bar{A}=\bartial UU^{-1}$ and $A=\partial VV^{-1}$
  and use the Polyakov-Wiegmann identity \cite{Polyakov:1984et}
\begin{equation}
  \mc{S}^G(gh)
  \ =\ \mc{S}^G(g)+\mc{S}^G(h)-\frac{i}{2\pi}\int
       \bigl\langle g^{-1}\bartial g,\partial hh^{-1}\bigr\rangle_G\,dz\wedge d\bar{z}
       \ \ ,
\end{equation}
  to write
\begin{equation}
  \label{eq:GaugedAction}
  \begin{split}
    \mc{S}^{G/H}(g,A,\bar{A})
    &\ =\ \mc{S}^G(g)+\frac{i}{2\pi}\int
          \Bigl\{-\bigl\langle\bar{A},A\bigr\rangle_H
                 +\bigl\langle\epsilon(\bar{A}),g\bar \epsilon(A)g^{-1}\bigr\rangle_G\\[2mm]
    &\hspace{3.5cm}+\bigl\langle\epsilon(\bar{A}),\partial gg^{-1}\bigr\rangle_G
           -\bigl\langle g^{-1}\bartial g,\bar \epsilon(A)\bigr\rangle_G\Bigr\}
           \,dz\wedge d\bar{z}\ \ .\raisetag{50pt}
  \end{split}
\end{equation}
  There are no terms which depend only on $U$ or $V$ in the previous
  action because the condition \eqref{eq:NoAnomaly} implies the relation
\begin{equation}
  \label{eq:NoAnomalyLagrangian}
   \mc{S}^G\bigl(\epsilon(h)\bigr)\ =\ \mc{S}^G\bigl(\bar \epsilon(h)\bigr)\ =\ \mc{S}^H(h)
  \quad\text{for all}\quad h\in H\ \ .
\end{equation}
  It is convenient to introduce the following compact notation for the
  Lagrangian \eqref{eq:GaugedAction}
\begin{equation}
  \label{eq:IntroMbb}
  \begin{split}
    \mc{S}^{G/H}(g,A,\bar{A})
    &\ =\ \mc{S}^G(g)+\frac{i}{2\pi}\int
          \Bigl\{\bar{A}^TMA+\bar{b}^TA+b^T\bar{A}\Bigr\}
          \,dz\wedge d\bar{z}\ \ ,
  \end{split}
\end{equation}
  where the gauge fields are expressed in coordinates with respect to
  some concrete basis of the Lie algebra and the matrix $M$ and the
  vectors $b$ and $\bar{b}$ are implicitly determined by comparing the
  integrands of \eqref{eq:GaugedAction} and \eqref{eq:IntroMbb}. The
  action is at most quadratic in the gauge fields and when the matrix
  $M$ is non-degenerate they can be easily integrated out. The
  resulting action is a $\s$-model whose metric and antisymmetric
  tensor can be inferred from
\begin{equation}
  \label{eq:CosetLagrangian}
  \begin{split}
    \mc{S}^{G/H}(g)
    &\ =\ \mc{S}^G(g)-\frac{i}{2\pi}\int
          \Bigl\{\,\bar{b}^TM^{-1}b\,\Bigr\}\,dz\wedge d\bar{z}\ \ .
  \end{split}
\end{equation}
  The background also includes a non-trivial
  dilaton~\cite{Witten:1991yr}, given up to constant terms by
\begin{equation} \label{eq:dilatonM}
  \Phi\ =  -\frac{1}{2}\ln\det M\ \ .
\end{equation}
  When the matrix $M$ is degenerate, the integration over the gauge
  fields results in the appearance of constraints for the $\s$-model
  fields. As we will see below in section \ref{subs:gaugefixing},
  the diagonal cosets in the class $(++)$ provide an example of this
  type. The occurrence of this and other somewhat unusual features is
  typical of gauged WZW models involving non-semisimple algebras
  \cite{Giveon:1995as}.

  For a more detailed discussion of asymmetric coset models in the
  bulk and on the boundary we refer the reader to
  \cite{Gawedzki:1989nj,Quella:2002fk}.

\subsection{The Heisenberg algebra and the associated group manifold}

  In this paper the general construction of the previous subsection
  will be applied to the diagonal cosets of the Heisenberg group
  $H_4$. The underlying Lie algebra, which will be denoted by the same
  symbol, is a four-dimensional non-semisimple Lie algebra. Its four
  generators $P_1, P_2, J$ and $K$ satisfy the following commutation
  relations
\begin{align}
  \label{eq:FinCommRel}
  [P_i,P_j]&\ =\ \epsilon_{ij}K\ ,&
  [J,P_i]&\ =\ \epsilon_{ij}P_j\ \ ,
\end{align}
  with $\epsilon_{12}=1$. In terms of the raising and lowering
  operators $P^\pm\ =\ P_1\pm iP_2$ the previous relations become
\begin{align}
  \label{eq:FinCommRel2}
  [P^+,P^-]&\ =\ - 2 i K\ ,&
  [J,P^\pm]&\ =\ \mp i P^\pm\ \ .
\end{align}
  In our conventions, the generators $P_1$ and $P_2$ are hermitian
  while $J$ and $K$ are anti-hermitian.

  The Heisenberg algebra admits a two-parameter family of invariant
  bilinear forms
\begin{align}
  \label{eq:metric-g}
  \langle P_i,P_j\rangle&\ =\ \Lambda \, \delta_{ij} \ \ , &
  \langle J,K\rangle&\ =\ \Lambda \ \ , & \langle J,J\rangle&\ =\ 2 \, \lambda \, \Lambda \ .
\end{align}
  By a rescaling of the generators and a redefinition $J\mapsto J-
  \lambda K$, it is always possible to set $\Lambda =1$ and $\lambda =
  0$ without affecting the commutation relations so that the metric assumes the standard form
\begin{align}
  \label{eq:metric}
  \langle P_i,P_j\rangle&\ =\  \delta_{ij} \ \ , &
  \langle J,K\rangle&\ =\ 1 \ .
\end{align}
  We also need an
  explicit parameterisation of the group elements. For the sake of easy
  comparison of our results with those of Antoniadis and
  Obers~\cite{Antoniadis:1994jx} we use
\begin{equation}
  g\ =\ e^{xP_1}\,e^{uJ}\,e^{yP_1}\,e^{vK} \ .
\end{equation}
  In this coordinate system the action of a single $H_4$ WZW model is
\be
  \mc{S}^{H_4}(g)
  \ =\ -\frac{i}{4 \pi} \int dz \wedge d \bar z \Bigl[
\p u \bar \p v + \p v \bar \p u +   \p x \bar \p x + \p y \bar \p y
+ 2 \cos u \p x \bar \p y  \Bigr] \ \ .
\ee
  In a similar way the group elements of $H_4 \times H_4$ will be
  parameterised by two sets of coordinates $(u_i, v_i, x_i, y_i)$, $i =
  1, 2$.

\subsection{Classification of diagonal coset models}

  In this section we shall provide a classification of all possible
  diagonal cosets of the Heisenberg group $H_4$. We will find three inequivalent families,
  each depending on six real parameters. As we will explain, for special choices of the
  parameters the models in two of these families coincide with the models studied in
  \cite{Antoniadis:1994jx}.

  Before we begin with a classification of all possible diagonal
  embeddings of the Heisenberg Lie algebra let us first define more
  precisely what we mean by a diagonal embedding. We recall that there
  is a canonical way to define the action of a Lie algebra $\g$ on a
  tensor product of two representations. The corresponding action is
  implemented in terms of the standard coproduct $\Delta$ which maps a
  generator $X\in\g$ to $\Delta(X)=X\otimes 1+1\otimes
  X\in\g\otimes\g$. Since the coproduct is injective and preserves the
  commutation relations it can be thought of as an embedding
  $\Delta:\g\to\g\oplus\g$. It is obvious that further embeddings can
  be obtained if one concatenates $\Delta$ with automorphisms of $\g$
  and $\g\oplus\g$, respectively. Inner automorphisms lead to
  equivalent embeddings so that one can focus on the group of outer
  automorphisms. If $n$ is the order of the group of outer automorphisms of
  $\g$, one should thus consider $2n^3$ a priori different
  possibilities, a factor $n$ for each of the Lie algebras $\g$ and a
  factor $2$ for the possibility of exchanging the two algebras in
  $\g\oplus\g$.\footnote{All the outer automorphisms occurring here
    and below should be thought of as representatives of equivalence
    classes of outer automorphisms modulo inner ones.}

  For simple Lie algebras the group of outer automorphisms is rather
  small, the maximum of $n=3$ being achieved for $\g=D_4$. On the
  other hand, the non-semisimple Lie algebra $\g=H_4$ offers a great
  variety of automorphisms. The existence of a two-parameter family of
  invariant metrics~\eqref{eq:metric-g} is paralleled by the existence
  of a continuous family of non-isometric automorphisms, which is the
  main reason for the significant number of diagonal coset models that
  can be constructed for $H_4$. In fact, the automorphisms of {\em
    simple} Lie algebras are always isometric. They just correspond to
  symmetries of the associated Dynkin diagram.

  In the case of $H_4$ it is possible to prove that the most general
  outer automorphism depends on a sign $\eta=\pm1$ and two continuous
  parameters $\mu\in\Real$ and  $\nu\in\Real_+$. Its action on the
  generators is the following,
\begin{equation}
  \label{eq:Auto}
  \Omega_\eta^{(\mu,\nu)}:\qquad(P_1,P_2,J,K)
  \ \mapsto\ \bigl(\nu P_1,\eta\nu P_2,\eta J+\mu K,\eta\nu^2K\bigr)
  \ \ .
\end{equation}
  These automorphisms are isometric only when $\nu = 1$ and $\m = 0$
  while in general one has
\begin{equation}
  \begin{split}
    \bigl\langle\Omega_\eta^{(\mu,\nu)}(J),\Omega_\eta^{(\mu,\nu)}(K)\bigr\rangle
    &\ =\ \nu^2\,\langle J,K\rangle \ , \\[2mm]
    \bigl\langle\Omega_\eta^{(\mu,\nu)}(J),\Omega_\eta^{(\mu,\nu)}(J)\bigr\rangle
    &\ =\ \langle J,J\rangle+2\mu\,\langle J,K\rangle\ \ .
  \end{split}
\end{equation}
Using the automorphisms $\Omega_\eta^{(\mu,\nu)}$ we can now
  construct the most general diagonal embedding $\epsilon:H_4\to
  H_4\oplus H_4$ which is given by
\begin{equation}
  \epsilon\ =\ \Omega_{\eta_1}^{(\mu_1,\nu_1)}\times\Omega_{\eta_2}^{(\mu_2,\nu_2)}
               \circ\Delta\circ\Omega_{\eta}^{(\mu,\nu)}\ \ .
               \label{eq:de-g}
\end{equation}
  In fact, it is easy to see that the automorphism
  $\Omega_{\eta}^{(\mu,\nu)}$ in \eqref{eq:de-g} is redundant since it
  can be removed by a redefinition of the parameters of the other two
  automorphisms. Accordingly, the most general diagonal embedding is
\begin{equation}
  \label{eq:Embedding}
  \begin{array}{rclcrcl}
    \epsilon(P_1)
    &\ =\ &\nu_1P_1^{(1)}+\nu_2P_1^{(2)}\ ,&\quad&
 \epsilon(J)
    &\ =\ &\eta_1J^{(1)}+\eta_2J^{(2)}+\mu_1K^{(1)}+\mu_2K^{(2)}  \ ,\\[2mm]
     \epsilon(P_2)
    &\ =\ &\eta_1\nu_1P_2^{(1)}+\eta_2\nu_2P_2^{(2)}\ ,&\quad&
    \epsilon(K)
    &\ =\ &\eta_1\nu_1^2K^{(1)}+\eta_2\nu_2^2K^{(2)} \ ,
  \end{array}
\end{equation}
  where the superscripts $(1)$ and $(2)$ refer to the two $H_4$
  factors. The embedding depends on four real
  parameters $\mu_1$, $\mu_2$, $\nu_1$ and $\nu_2$ and on the pair of
  signs $(\eta_1\eta_2)$ that will subsequently be called the class of
  the embedding.

  We now turn to the classification of possible diagonal coset
  models. We first note that using the outer automorphisms just
  described we can always choose on $H_4\times H_4$ the standard
  invariant bilinear form  with  $\Lambda=1$ and $\lambda=0$ for both
  factors. This simply amounts to a reparameterisation of the group
  elements.

  We can then choose two  embeddings of the form~\eqref{eq:Embedding}
  and require that they satisfy the constraint~\eqref{eq:NoAnomaly},
  which ensures that we are gauging an anomaly-free subgroup. In terms
  of the parameters $(\eta_i,\mu_i,\nu_i)$ and
  $(\etabb_i,\mubb_i,\nubb_i)$ of the two embeddings,
  Eq.~\eqref{eq:NoAnomaly} becomes
\begin{align}
  \label{eq:Constraint}
  \nu_1^2+\nu_2^2&\ =\ \nubb_1^2+\nubb_2^2\ \equiv \ \Lambda\ \ ,&
  \eta_1\mu_1+\eta_2\mu_2&\ =\ \etabb_1\mubb_1+\etabb_2\mubb_2
  \ \equiv \ \Lambda\lambda \ .
\end{align}
  The left part of both equations should be read as a consistency
  condition while the right part defines the constants $\Lambda$ and
  $\lambda$. It is convenient to write the general solution of the
  previous constraint equations in the following form
\begin{align}
  (\nu_1,\nu_2)
  &\ =\ \sqrt{\Lambda}\,(\cos\alpha,\sin\alpha)\ \ ,&
  (\nubb_1,\nubb_2)
  &\ =\ \sqrt{\Lambda}\,(\cos\balpha,\sin\balpha)\ \ ,\nonumber\\[2mm]
  (\mu_1,\mu_2)
  &\ =\ \bigl(\eta_1\Lambda\mu,\eta_2\Lambda(\lambda-\mu)\bigr)\ \ ,&
  (\mubb_1,\mubb_2)
  &\ =\ \bigl(\etabb_1\Lambda\mubb,\etabb_2\Lambda(\lambda-\mubb)\bigr)
  \ \ ,
\end{align}
  with $\alpha, \bar \alpha \in (0, \pi/ 2)$. At this point it seems
  that we are left with four discrete and six continuous
  parameters. However two of the discrete parameters can be removed
  using the freedom of reparameterising the group elements. We already
  used this freedom to choose the standard metric on $H_4\oplus H_4$
  but we can still act with the outer isometric automorphism
  $\Omega_{-1}^{(0,1)}$. In this way we can set for instance $\eta_1 =
  \eta_2=1$.

  We thus arrive at the conclusion that the diagonal coset models
  based on the Heisenberg group $H_4$ are specified by two discrete
  parameters and six continuous parameters. Since the physical
  properties will strongly depend on the particular choice of signs
  $(\etabb_1\etabb_2)$, this tuple will be called the class of the
  coset. It labels distinct families of models. In each family there
  are two special choices of parameter, namely $\balpha=\alpha$ and
  $\balpha=\pi/2-\alpha$. The resulting models will be referred to as
  ``symmetric'' and ``twisted'' gaugings respectively. The $(++)$ and
  $(--)$ families of diagonal cosets with $\alpha = \bar \alpha$ and
  $\lambda = \m = \bar \m = 0$ correspond respectively to the vector
  gauged model and to the vector-axial gauged model studied in
  \cite{Antoniadis:1994jx}.

  From the explicit construction that will be performed in the next
  section, it turns out that the parameters $\alpha$ and $\balpha$
  always label inequivalent models. On the other hand the parameter
  $\Lambda$ disappears from the action. For the families $(--)$ and
  $(+-)$, the parameters $\mu$, $\mubb$ and $\lambda$ can be removed
  by a simple coordinate redefinition and do not generically affect
  the spectrum of the model, unless the coordinates $v_1$ or $v_2$ are
  compact. Finally, the family $(++)$ has a non-trivial dependence on
  the difference $\m - \bar \m$ which can however be removed by a
  T-duality transformation. Consequently, the number of physical
  parameters is slightly smaller than indicated by our purely
  algebraic reasoning.

\subsection{\label{subs:gaugefixing}Derivation of coset data and gauge fixing}

  In this section we compute the quantities $M$, $b$ and $\bar{b}$
  that were defined in Eq.~\eqref{eq:IntroMbb}. Thereby we completely
  specify the action of the gauged WZW model. We also describe our
  gauge choices for the different classes of models. The gauge field
  takes values in the Lie algebra $H_4$ and can be written as
\begin{equation}
  A\ =\ A_1P_1+A_2P_2+A_3J+A_4K \ ,
  \qquad
  \bar{A}\ =\ \bar{A}_1P_1+\bar{A}_2P_2+\bar{A}_3J+\bar{A}_4K\ \ .
\end{equation}
  We embed these fields in the numerator algebra using the two
  embeddings $\e$ and $\bar \e$,
\ba
  \hspace{-0.8cm}
  \epsilon(\bar{A})
     &=& \nu_1\bar{A}_1\,P_1^{(1)}+\nu_1\bar{A}_2\,P_2^{(1)}
          +\bar{A}_3\,J^{(1)}+\bigl[\nu_1^2\bar{A}_4+\mu_1\bar{A}_3\bigr]K^{(1)}
          +(1\leftrightarrow2)\ \ , \nb \\
   \hspace{-0.8cm} \bar \epsilon(A)
    &=& \nubb_1A_1\,P_1^{(1)}+\etabb_1\nubb_1A_2\,P_2^{(1)}
          +\etabb_1A_3J^{(1)}+\bigl[\etabb_1\nubb_1^2A_4+\mubb_1A_3\bigr]K^{(1)}
          +(1\leftrightarrow2)\ \ .
\ea
  The expression $(1\leftrightarrow2)$ indicates the presence of a
  similar contribution in the second factor, obtained by replacing the
  label $1$ by the label $2$ in the superscripts of the generators and
  in the subscripts of the parameters. The explicit form of the
  matrices $M$, $b$ and $\bar{b}$ is then
\begin{equation}
  \label{eq:CosetData}
  \begin{split}
    M&\ =\ \smat
             -\frac{\Lambda}{2}+\nu_1\nubb_1c_1 & -\etabb_1\nu_1\nubb_1s_1 &
             \etabb_1\nu_1y_1s_1 & 0 \\
             \nu_1\nubb_1s_1 & -\frac{\Lambda}{2}+\etabb_1\nu_1\nubb_1c_1 &
             -\etabb_1\nu_1(x_1+y_1c_1) & 0 \\
             \nubb_1x_1s_1 & \etabb_1\nubb_1(x_1c_1+y_1) &
             -\frac{\Lambda\lambda}{2}+\etabb_1\mu_1+\mubb_1-\frac{1}{2}\etabb_1(x_1^2+y_1^2+2x_1y_1c_1)
             & -\frac{\Lambda}{2}+\etabb_1\nubb_1^2 \\
             0 & 0 & -\frac{\Lambda}{2}+\etabb_1\nu_1^2 & 0
           \stam
            +(1\leftrightarrow2)\\[2mm]
    b&\ =\ \smat\nu_1(\partial x_1+c_1\partial y_1)\\[2mm]
           \nu_1(s_1\partial y_1-x_1\partial u_1)\\[2mm]
           \partial v_1+x_1s_1\partial y_1-\frac{x_1^2}{2}\partial u_1+\mu_1\partial u_1\\[2mm]
           \nu_1^2\partial u_1\stam
           +(1\leftrightarrow2)\\[2mm]
    \bar{b}&\ =\ \smat-\nubb_1(c_1\bartial x_1+\bartial y_1)\\
           -\etabb_1\nubb_1(-s_1\bartial x_1+y_1\bartial u_1)\\[2mm]
           -\mubb_1\bartial u_1-\etabb_1\bigl(\bartial v_1+y_1 s_1\bartial x_1-\frac{y_1^2}{2}\bartial u_1\bigr)\\[2mm]
           -\etabb_1\nubb_1^2\bartial u_1\stam
           +(1\leftrightarrow2)\ \ .\raisetag{108pt}
  \end{split}
\end{equation}
 In the previous expressions we introduced the abbreviations
  $c_i=\cos(u_i)$ and $s_i=\sin(u_i)$. As before, the notation
  $(1\leftrightarrow2)$ stands for an additional term
  identical to the first except for the relabelling of all the
  indices.

  We now turn to the gauge fixing
  conditions for the local symmetry
\be
  g\mapsto\epsilon(h)g\bar{\epsilon}(h^{-1}) \ , \hspace{0,8cm}
  A \mapsto h ( A - \p) h^{-1} \ , \hspace{0,8cm}
  \bar A \mapsto h (  \bar A -  \bar \p) h^{-1}
\ee
  of our models.
  For the two classes $(--)$ and $(+-)$ we choose a gauge
  where the $\s$-model fields satisfy
  the following relations,
\begin{align}
  \label{eq:GeneralGaugeFixing}
  x_1&\ =\ x\cos\alpha\ ,&
  y_1&\ =\ y\cos\bar \alpha\ ,&
  u_1&\ =\ u\ ,&
  v_1&\ =\ v\nonumber\\[2mm]
  x_2&\ =\ x\sin\alpha\ ,&
  y_2&\ =\ - y\sin\bar \alpha\ ,&
  u_2&\ =\ -u\ ,&
  v_2&\ =\ -v\ \ .
\end{align}
  The resulting matrix $M$ is non-singular and after integrating over the gauge
  fields one obtains a $\s$-model action of the form displayed in
  Eq.~\eqref{eq:CosetLagrangian} with a dilaton given by
  Eq.~\eqref{eq:dilatonM}. The resulting background fields are
  discussed in sections~\ref{sc:CosetsMM} and~\ref{sc:CosetsPM}.

  For the cosets of type $(++)$ the matrix $M$ turns out to be
  singular and it has the following form \be M = \smat \ \tilde{M} & \
  0 \ \\[2mm] 0 & \ 0 \ \stam \ , \ee where $\tilde{M}$ is a
  non-singular $3 \times 3$ matrix. We first fix the gauge freedom
  associated with the transformations generated by $P_1$, $P_2$ and
  $J$ setting
\begin{equation}
  \label{eq:GaugeFixingPP}
  x_1\ =\ x\cos\alpha\ ,
  \qquad
  x_2\ =\ x\sin\alpha\ ,
  \qquad
  y_1 = y_2 = 0 \ \ .
\end{equation}
  Since in this case the fourth row and column of the matrix $M$
  vanish, the fields $A_4$ and $\bar{A}_4$ appear in the Lagrangian of
  the gauged WZW model only in the following two linear terms
\begin{equation}
  {\cal L}_{A_4} =
2 \bigl(\nu_1^2\partial u_1+\nu_2^2\partial u_2\bigr)\bar{A}_4\ - 2
\bigl(\nubb_1^2\bartial u_1+\nubb_2^2\bartial u_2\bigr)A_4 \ \ .
\end{equation}
Up to a total derivative the previous expression is equivalent to
\begin{equation}
  {\cal L}_{A_4} = (\p \bar A_4 - \bar \p A_4)  \, U + (\p \bar A_4 + \bar \p A_4) \, V \ \ ,
\end{equation}
where \be
  U \equiv (\nu_1^2
  + \nubb_1^2)u_1 + (\nu_2^2 + \nubb_2^2)u_2 \ , \hspace{0.8cm} V
  \equiv (\nu_1^2 - \nubb_1^2)u_1 + (\nu_2^2 - \nubb_2^2)u_2 \ \ .
\ee
  We can fix the gauge freedom associated with the transformations
  generated by $K$ choosing the gauge $\p \bar A_4 + \bar \p A_4 =
  0$. The integration over the gauge field $A_4$ then leads to the
  constraint $U = 0$. We can modify this constraint by adding the
  total derivative ${\cal L}_\rho = - \Lambda\rho (\p \bar A_4 - \bar
  \p A_4) $ to the Lagrangian where $\rho$ is a real constant. The
  constraint then becomes $U = \Lambda\rho$ and its general solution
  reads
\begin{equation}
  u_1\ =\ (1 - \gamma)u + \rho\ \ ,
  \hspace{1cm} u_2 = -(1+\gamma) u + \rho\ \ ,
\end{equation}
  where
\begin{equation}
  \gamma\ =\ \cos(\alpha - \bar \alpha)\cos(\alpha + \bar \alpha) \ \ .
\end{equation}
Due to the presence of the constraint $U = \rho$, the construction
of the asymmetric models with $\n_i \ne \bar \n_i$ might not be
entirely straightforward. In particular it is not clear if it leads
to non-trivial conformally invariant $\s$-models. We leave the
general discussion to a future publication and in the rest of this
paper we shall only consider models in the $(++)$ class with $\n_i =
\bar \n_i$ or, equivalently, $\alpha = \bar \alpha$. Since the matrix $\tilde{M}$
is non-singular, after integrating over the remaining gauge fields
one obtains again a $\s$-model action of the form displayed in
  Eq.~\eqref{eq:CosetLagrangian} with a dilaton given by
  Eq.~\eqref{eq:dilatonM}. The resulting background fields are discussed
   in section \ref{sc:CosetsPP}.

\section{\label{sc:CosetGeometries}Coset geometries}

  The coset construction for the various embeddings considered in the
  previous section gives rise to a number of interesting geometries.
  They are all particular examples of a class of string backgrounds
  related to plane waves by abelian T-duality transformations
  \cite{Klimcik:1993kd}. For this reason we
  briefly review the properties of gravitational plane waves at the
  beginning of this section. We then discuss the geometries associated
  with our three families of diagonal cosets. The general features of
  each class of models will be illustrated by two simple choices of
  parameters: the symmetric gauging $\alpha = \bar \alpha$ and the
  twisted gauging $\alpha = \pi/2 - \bar \alpha$. As we will see the
  symmetric gaugings always lead to singular backgrounds while the
  twisted gaugings are non-singular for generic values of the
  parameters. Consequently the asymmetry in the parameters $\alpha$
  and $\bar \alpha$ allows to interpolate between singular and
  non-singular backgrounds. The parameters $\lambda$, $\m$ and $\bar
  \m$ can generically be removed by a change of coordinates.

  The background fields that correspond to the most general choices of
  parameters are collected in appendix~\ref{ap:FullGeometries} since
  the resulting expressions are quite lengthy.

\subsection{General plane waves}

  The curved space-times that correspond to the diagonal cosets of the
  Heisenberg group belong to a class of four-dimensional string
  backgrounds introduced in \cite{Klimcik:1993kd}. These models have a
  covariantly constant null Killing vector and are either plane waves
  or are related to plane waves by a T-duality transformation with
  respect to an abelian non-null isometry.  Following the notation in
  \cite{Antoniadis:1994jx}, the metric, dilaton and antisymmetric
  tensor of a general plane wave string background are given by
\ba
  \label{eq:StdPW}
  ds^2
  \ &=& \ 2d\zeta d\zetab
       -2\Bigl[f(u)\zeta^2+\bar{f}(u)\zetab^2+F(u)\zeta\zetab\Bigr]du^2
       -2dudv\ \ . \nb \\[2mm]
B_{\zeta \zetab} \, &=& \, i b(u) \ , \hspace{1cm} \F = -\log g(u)
\ \ .
\ea
  Here $f(u)$ is a complex function, $g(u)$ and $b(u)$ are real
  functions and the function $F(u)$ is given by
\be
  F(u)\ =\ -\p^2_u \log g(u) + \frac{1}{4}\bigl[ \p_u b(u) \bigr]^2 \ .
  \label{eq:betafunction}
\ee
  The previous equation is equivalent to the condition $\beta^G = 0$
  where $\beta^G$ is the one-loop beta function for the metric of the
  $\s$-model. In fact the only non-trivial component of the Ricci
  tensor for the metric in \eqref{eq:StdPW} is
\begin{equation}
  \label{eq:RF}
  R_{uu}\ =\ 2F(u)\ \ .
\end{equation}
  The one-loop beta functions $\beta^\F$ and $\beta^{B}$ for the
  dilaton and the antisymmetric tensor vanish identically.

  The backgrounds in \eqref{eq:StdPW} were classified in
  \cite{Antoniadis:1994jx} according to their isometries. For generic
  choices of the functions $f(u)$ and $F(u)$ they have five
  isometries. There is an additional isometry when $f(u)=0$ or when
  $f(u)$ and $F(u)$ are both constant. Finally when $f(u)=0$ and
  $F(u)$ is constant the background \eqref{eq:StdPW} has seven
  isometries and coincides with the $H_4$ WZW model
  \cite{Nappi:1993ie}, the maximally symmetric plane wave in four
  dimensions.

\subsection{\label{sc:CosetsMM}The cosets of type $(--)$}

  We can now follow the procedure described in the previous section
  and derive the background fields of the coset models that belong to
  the $(--)$ class. The symmetric gauging coincides with the
  axial-vector gauging discussed in \cite{Antoniadis:1994jx} and leads
  to singular plane waves. The general models with $\alpha\neq\balpha$
  are still plane waves but now without singularities. In fact from
  the explicit form of the background fields of the twisted gaugings
  one can easily see that the corresponding models smoothly
  interpolate between non-singular spaces and a special singular point
  where the symmetric and twisted gauging coincide. For readers
  interested in the general asymmetric coset of type $(--)$ we
  collected the corresponding background fields in
  appendix~\ref{ap:FullGeometries}.

\subsubsection{Symmetric gauging}

  In order to make contact with the work of Antoniadis and Obers
  \cite{Antoniadis:1994jx} it is useful to replace the angle
  $\alpha=\balpha$ by the new parameter $\nu=\tan\alpha$. The
  parameters of the model are then specified in the following way
\begin{align}
  \nu_1&\ =\ 1&
  \nu_2&\ =\ \nu&
  \mu_1&\ =\ \Lambda\mu&
  \mu_2&\ =\ \Lambda(\lambda-\mu)\nonumber\\[2mm]
  \bar{\nu}_1&\ =\ 1&
  \bar{\nu}_2&\ =\ \nu&
  \mubb_1&\ =\ -\Lambda\mubb&
  \mubb_2&\ =\ -\Lambda(\lambda-\mubb)\ \ ,
\end{align}
  with $\Lambda=1+\nu^2$. Given this choice of parameters and the gauge fixing
  \eqref{eq:GeneralGaugeFixing}, we obtain the following metric
\begin{equation}
  \label{eq:MMSymmetric}
  \begin{split}
    ds^2
    &\ =\ 4dudv+\frac{r^+(u)}{2\nu^2\bigl(1-\cos(u)\bigr)}\,dx^2
          +\frac{1-\nu^4}{\nu^2\bigl(1-\cos(u)\bigr)}\,dxdy
          +\frac{r^-(u)}{2\nu^2\bigl(1-\cos(u)\bigr)}\,dy^2\\[2mm]
    &\qquad-\frac{(1-\nu^2)\bigl[(1-\nu^4)x+(4\nu^2+r^+(u))y\bigr]}{2\nu^2(1+\nu^2)\sin(u)}\,dxdu\\[2mm]
    &\qquad-\frac{(1-\nu^4)r^-(u)x+\bigl((1+\nu^2)^4-16\nu^4\cos(u)\bigr)y}{2\nu^2(1+\nu^2)^2\sin(u)}\,dydu\\[2mm]
    &\qquad+\frac{(1-\nu^2)^2r^-(u)x^2}{8\nu^2(1+\nu^2)^2\bigl(1+\cos(u)\bigr)}\,du^2
           +\frac{(1-\nu^4)xy}{4\nu^2\bigl(1+\cos(u)\bigr)}\,du^2\\[2mm]
    &\qquad+\frac{(1-\nu^2)^2\bigl(8\nu^2+r^+(u)\bigr)y^2}{8\nu^2(1+\nu^2)^2\bigl(1+\cos(u)\bigr)}\,du^2
          +\frac{2(1-\nu^2)\bigl[2\lambda-(1+\nu^2)(\mu+\mubb)\bigr]}{1+\nu^2}\,du^2\ \
          ,
  \end{split}
\end{equation}
  where
\begin{equation}
  \label{eq:SmallR}
  r^\pm(u)\ =\ 1+\nu^4\pm2\nu^2\cos(u)\ \ .
\end{equation}
  The last term in the metric can be removed by the
  coordinate transformation $v \to v - \xi u$ with constant
  $\xi$. Since the three-form flux $H$ vanishes, the only other
  non-trivial background field is the dilaton
\begin{equation}
  \Phi\ =\ -\frac{1}{2}\ln\sin^2(u)\ \ .
\end{equation}
  It is easy to check that the metric above coincides with the one in
  \cite{Antoniadis:1994jx} after setting
  $\nu=\sqrt{(1-\kappa)/(1+\kappa)}$ and rescaling the coordinate
  $v$. The resulting background is a singular plane wave belonging to
  the family \eqref{eq:StdPW}. The only non-vanishing component of the
  Ricci tensor is $R_{uu}$ and it turns out to be proportional to
\begin{equation}
  F(u)\ =\ -1/\sin^2(u)\ \ .
\end{equation}
  Following the series of coordinate transformations that have been
  described in \cite{Antoniadis:1994jx} one can also determine the
  function
\begin{equation}
  f(u)\ =\
  \frac{1}{2\sin^2(u)}\bigl[\cos(u)-i\kappa\sin(u)\bigr]\,e^{i\kappa u}\ \
  ,
\end{equation}
  which together with $g(u) = \sin u$ and $b(u) = 0$ completely specifies
  the background in \eqref{eq:StdPW}.

\subsubsection{Twisted gauging}

  As mentioned above, all the other models in the $(--)$ class
  of diagonal cosets are non-singular. Their main features are
  well illustrated by the following simple choice of parameters
\begin{align}
  \nu_1&\ =\ 1&
  \nu_2&\ =\ \nu&
  \mu_1&\ =\ \Lambda\mu&
  \mu_2&\ =\ \Lambda(\lambda-\mu)\nonumber\\[2mm]
  \nubb_1&\ =\ \nu&
  \nubb_2&\ =\ 1&
  \mubb_1&\ =\ -\Lambda\mubb&
  \mubb_2&\ =\ -\Lambda(\lambda-\mubb) \ ,
\end{align}
  with $\Lambda=1+\nu^2$. Note that in this case we have $\n_1 = \bar \n_2$ and
  $\n_2 = \bar \n_1$ which is the reason for the name ``twisted
  gauging'' given to this class of models. With the previous choice of parameters
  and the gauge fixing
  \eqref{eq:GeneralGaugeFixing} the metric reads
\begin{equation}
  \label{eq:MMTwisted}
  \begin{split}
    ds^2
    &\ =\ 4dudv+\frac{R^+(u)}{R^-(u)}\,dx^2
          -\frac{2(1-\nu^2)}{R^-(u)}\,dxdy
          +dy^2
          -\frac{2\nu(1-\nu^2)\sin(u)}{(1+\nu^2)R^-(u)}\,ydxdu\\[2mm]
    &\qquad+\frac{2\nu(1-\nu^2)\sin(u)}{(1+\nu^2)R^+(u)}\,xdydu
          -\frac{8\nu^2\bigl(2\nu-(1+\nu^2)\cos(u)\bigr)\sin(u)}{(1+\nu^2)R^+(u)R^-(u)}\,
          ydydu\\[2mm]
    &\qquad-\frac{(1-\nu^2)^2R^-(u)x^2-2(1+5\nu^2-5\nu^4-\nu^6)xy+(1-\nu^2)^2R^+(u)y^2}
    {4(1+\nu^2)^2R^+(u)}\,du^2\\[2mm]
    &\qquad-\frac{2(1-\nu^2)^2\lambda-2(1-\nu^4)(\mu-\mubb)}{1+\nu^2}\,du^2
    \ \ ,
  \end{split}
\end{equation}
 where
\begin{equation}
  \label{eq:CapR}
  R^\pm(u)\ =\ 1+\nu^2\pm2\nu\cos u\ \ .
\end{equation}
The dilaton is given by
\begin{equation}
  \Phi\ =\ -\frac{1}{2}\ln\bigl(1+\nu^4-2\nu^2\cos(2u)\bigr)
      \ =\ -\frac{1}{2}\ln\bigl(R^+(u)R^-(u)\bigr)\ \ ,
\end{equation}
and in this case there is also  a non-trivial
  two-form field $B_{\m\n}$ with flux

\begin{equation}
  H\ =\ \frac{2\nu(1-\nu^2)\bigl(1+4\nu^2+\nu^4+2\nu^2\cos(2u)\bigr)\sin(u)}
  {(1+\nu^2)R^+(u)R^-(u)^2}\,dx\wedge dy\wedge du\ \ .
\end{equation}

The previous background fields describe a non-singular plane wave.
In fact the $R_{uu}$ component of the Ricci tensor is
\begin{equation}
  \begin{split}
    F(u)
    &\ =\ \frac{1}{4(1+\nu^2)^2\bigl(R^+(u)R^-(u)\bigr)^2}
          \Bigl[(1+\nu^2)^4(1-18\nu^2+\nu^4)\\[2mm]
    &\qquad\quad+8\nu^2(1+\nu^2)^2(5-2\nu^2+5\nu^4)
          \cos^2(u)+16\nu^4(1-\nu^2)^2\cos^4(u)\Bigr]\ \ ,
  \end{split}
\end{equation}
  Like before it is also possible to determine the function $f(u)$
  specifying the background \eqref{eq:StdPW}. One finds
\begin{equation}
  f(u)\ =\ -\frac{4 \nu^3 \cos(u) }{(1+\nu^2)R^+(u)R^-(u)}\ \ .
\end{equation}
  In contrast to the symmetric case this time $f(u)$ does not exhibit
  an imaginary part. Note that both quantities $F(u)$ and $f(u)$ are
  always regular for $\nu\neq1$. Accordingly, by deforming the action
  of the subgroup from the symmetric to the twisted action we
  completely removed all singularities.

\subsection{\label{sc:CosetsPM}The cosets of type $(+-)$}

  The second class of cosets corresponds to $\etabb_1=1$ and
  $\etabb_2=-1$.  As before we shall discuss two
  special choices of parameters, the symmetric and the twisted
  gaugings. The geometric data for the general choice of parameters
  are displayed in appendix~\ref{ap:FullGeometries}.

\subsubsection{Symmetric gauging}

  We start our discussion with the symmetric case
\begin{align}
  \nu_1&\ =\ 1&
  \nu_2&\ =\ \nu&
  \mu_1&\ =\ \Lambda\mu&
  \mu_2&\ =\ \Lambda(\lambda-\mu)\nonumber\\[2mm]
  \bar{\nu}_1&\ =\ 1&
  \bar{\nu }_2&\ =\ \nu&
  \mubb_1&\ =\ \Lambda\mubb&
  \mubb_2&\ =\ -\Lambda(\lambda-\mubb)\ \ ,
\end{align}
 with $\Lambda=1+\nu^2$. Following the standard procedure we obtain the metric
\begin{equation}
  \begin{split}
    ds^2
    &\ =\ 2(1+\nu^2)/\nu^2dudv
          +\frac{1+\cos(u)}{1-\cos(u)}\,\nu^2dx^2
          +\frac{2\nu^2\bigl(1+\nu^2-2\cos(u)\bigr)\bigl(1+\cos(u)\bigr)}{(1+\nu^2)(1-\cos(u))}\,dxdy\\[2mm]
    &\qquad+\frac{\nu^2\bigl[(1+\nu^2)^2-(3+2\nu^2-\nu^4)\cos(u)+4(1-\nu^2)\cos^2(u)\bigr]}
    {(1+\nu^2)^2\bigl(1-\cos(u)\bigr)}\,dy^2\\[2mm]
    &\qquad
          +(1-\nu^4)\cot(u/2)/\nu^2\,xdxdu\\[2mm]
    &\qquad+(1-\nu^2)\bigl[1-\nu^4+2\nu^2\bigl(1+\cos(u)\bigr)\bigr]\cot(u/2)/\nu^2(1+\nu^2)\,ydxdu\\[2mm]
    &\qquad+\bigl(1+\nu^4-2\nu^2\cos(u)\bigr)\cot(u/2)/\nu^2\,xdydu\\[2mm]
    &\qquad+\bigl(1+2\nu^2+4\nu^4+2\nu^6-\nu^8-4\nu^4(3-\nu^2)\cos(u)\bigr)\cot(u/2)/\nu^2(1+\nu^2)^2\,ydydu\\[2mm]
    &\qquad-(1-\nu^4)/4\nu^2\,x^2du^2
           -(1-\nu^2)\bigl(3-\nu^2+2\cos(u)\bigr)/2\nu^2\,xydu^2\\[2mm]
    &\qquad-\frac{(1-\nu^2)\bigl(1+11\nu^2-5\nu^4+\nu^6+4(1+4\nu^2-\nu^4)\cos(u)\bigr)}
    {4\nu^2(1+\nu^2)^2}\,y^2du^2\\[2mm]
    &\qquad+(1+\nu^2)\bigl(-2\lambda+\mu(1+\nu^2)^2+\mubb(1-\nu^4)\bigr)/\nu^4\,du^2\ \ .
  \end{split}
\end{equation}
  The last term can be removed by the coordinate transformation
  $v \rightarrow v - \xi u$ with constant $\xi$. The B-field is
  pure gauge and the dilaton is given by
\begin{equation}
  \Phi\ =\ -\frac{1}{2}\ln\sin^2\frac{u}{2}\ \ .
\end{equation}
   The $R_{uu}$ component of the Ricci tensor is
\begin{equation}
  F(u)\ =\ -1/4\sin^2\frac{u}{2}\ \ ,
\end{equation}
and therefore we obtain again a singular plane wave background.

\subsubsection{Twisted gauging}

Also in this case we illustrate the main features of the models that
correspond to a general choice of parameters with the twisted
gaugings
\begin{align}
  \nu_1&\ =\ 1&
  \nu_2&\ =\ \nu&
  \mu_1&\ =\ \Lambda\mu&
  \mu_2&\ =\ \Lambda(\lambda-\mu)\nonumber\\[2mm]
  \bar{\nu }_1&\ =\ \nu&
  \bar{\nu }_2&\ =\ 1&
  \mubb_1&\ =\ \Lambda\mubb&
  \mubb_2&\ =\ -\Lambda(\lambda-\mubb) \ ,
\end{align}
  with $\Lambda=1+\nu^2$. The metric reads
\begin{equation}
  \begin{split}
    ds^2
    &\ =\ 2(1+\nu^2)/\nu^2\,dudv+\frac{R^+(u)}{R^-(u)}\,dx^2
          -\frac{2\bigl(1+4\nu^2-\nu^4-4\nu^2\cos^2(u)\bigr)}{(1+\nu^2)\,R^-(u)}\,dxdy\\[2mm]
    &\qquad+\frac{1+11\nu^2-5\nu^4+\nu^6-2\nu(1+\nu^2)^2\cos(u)-8\nu^2(1-\nu^2)\cos^2(u)}
    {(1+\nu^2)^2\,R^-(u)}\,dy^2\\[2mm]
    &\qquad+\frac{2(1-\nu^2)\sin(u)}{\nu\,R^-(u)}\,xdxdu
          +\frac{2(1-\nu^2)\sin(2u)}{(1+\nu^2)\,R^-(u)}\,ydxdu\\[2mm]
    &\qquad-\frac{\bigl[2(1-3\nu^2-\nu^4-\nu^6)+4\nu^3(1+\nu^2)\cos(u)\bigr]\sin(u)}
    {\nu(1+\nu^2)\,R^-(u)}\,xdydu\\[2mm]
    &\qquad-\frac{\bigl[2\nu(1-3\nu^2-5\nu^4-\nu^6)+4(1-3\nu^2+5\nu^4+\nu^6)\cos(u)\bigr]
    \sin(u)}{(1+\nu^2)^2\,R^-(u)}\,ydydu\\[2mm]
    &\qquad-\frac{1-\nu^4}{4\nu^2}\,x^2du^2
           +\frac{1-5\nu^2+3\nu^4+\nu^6-2\nu(1-\nu^4)\cos(u)}{2\nu^2(1+\nu^2)}\,xydu^2\\[2mm]
    &\qquad-\frac{(1-\nu^2)\bigl[(1+\nu^2)^3-4\nu(1-4\nu^2-\nu^4)\cos(u)\bigr]}
    {4\nu^2(1+\nu^2)^2}\,y^2du^2\\[2mm]
    &\qquad+(1+\nu^2)\bigl[\mubb(1-\nu^4)-\lambda(1+2\nu^2-\nu^4)+\mu(1+\nu^2)^2\bigr]/\nu^2\,du^2
    \ \ ,
  \end{split}
\end{equation}
  where the functions $R^\pm(u)$ were defined in
  \eqref{eq:CapR}. The background also supports a dilaton
\begin{equation}
  \Phi\ =\ -\frac{1}{2}\ln\bigl(R^-(u)\bigr) \ .
\end{equation}
{\noindent} and a
  non-trivial three-form flux
\begin{equation}
  H\ =\ \frac{(1-\nu^2)\bigl(1-\nu^2+2\nu\cos(u)\bigr)
  \bigl(1+4\nu^2+\nu^4-2\nu(1+\nu^2)\cos(u)\bigr)\sin(u)}{\nu(1+\nu^2)R^-(u)^2}\,dx\wedge dy\wedge du\ \
  .
\end{equation}

{\noindent} The $R_{uu}$ component of the Ricci tensor is given by
\begin{align}
  F(u)
  &\ = \ \frac{1}{16\nu^4\,R^-(u)^2}\Bigl[1+6\nu^2+3\nu^4-52\nu^6+3\nu^8+6\nu^{10}+\nu^{12}\\[2mm]
  & \qquad-4\nu(1+\nu^2)(1+2\nu^2-10\nu^4+2\nu^6+\nu^8)\cos(u)+4\nu^2(1-\nu^4)^2\cos^2(u)\Bigr]\ \nb \
  .
\end{align}
  From the previous expression we can
  see that also for this class of models the singular behaviour of the
  background fields of the symmetric coset is regularised by an
  asymmetric choice of parameters.

\subsection{\label{sc:CosetsPP}The cosets of type $(++)$}

The last family of models corresponds to the choice $\etabb_1 = \etabb_2
= 1$. In this case, as mentioned in section \ref{sc:Data}, we
restrict our analysis to the symmetric models with $\alpha=\bar
\alpha$, due to potential subtleties with the constraints and the
gauge fixing procedure for the general models. With this choice of parameters
the metric, the antisymmetric tensor and the dilaton are
\begin{equation}
  \label{eq:MetricPP}
  \begin{split}
    ds^2
    &\ =\ \frac{1}{\Delta(u)}\left[
            F_+(u)dx^2 + \frac{4l_2(u)}{x} dx d\f +\frac{F_-(u)}{x^2} d \f^2
            -\frac{4\Lambda^2(\m-\bar \mu)^2}{x^2} F_-(u) du^2 \right ] - 2 du dv \ \ ,\\[2mm]
    &\qquad\qquad\quad B_{u\f}
    \ =\ \frac{\Lambda (\m-\bar \m)}{x^2} \frac{2 F_-(u)}{\Delta(u)}  \ \ ,
    \hspace{2cm}
    \Phi\ =\ - \frac{1}{2} \ln\bigl[x^2 \Delta(u)\bigr] \ \ ,\raisetag{20pt}
  \end{split}
\end{equation}
  where we defined the auxiliary functions
\begin{equation}
  \Delta(u)
  \ =\ \sin^2(2\alpha)\,\sin^2(u) \ ,
  \qquad
  F_{\pm}(u)
  \ =\ 1 + l_1^2(u) + l_2^2(u) \pm 2 l_1(u) \ \ ,
\end{equation}
  with
\begin{equation}
  l_1(u)\ =\ \cos u \cos(\gamma u - \rho) + \gamma \sin u \sin(\gamma u - \rho)
      \ ,
\end{equation}
\begin{equation}
  l_2(u)\ =\ \gamma \sin u \cos( \gamma u - \rho) - \cos u \sin(\gamma u - \rho)
       \ ,
\end{equation}
and $\gamma = \cos(2\alpha)$. In the process of recovering the
background data from the gauged WZW Lagrangian
\eqref{eq:CosetLagrangian} we introduced the new coordinates
\begin{equation}
  \f\ =\ \frac{v_1 + v_2}{2} \ \ ,
  \hspace{1cm}
  v\ =\ -(1-\gamma)v_1 + (1+\gamma)v_2 \ \ ,
\end{equation}
  and performed the following change of variables
\begin{equation}
  \f \mapsto \f + 2\Lambda\bigl[\lambda(1+\gamma) -(\mu+\mubb)\bigr] u
  \ .
\end{equation}
  When $\mu=\mubb=0$ the curved backgrounds \eqref{eq:MetricPP} coincide with the
  vector-gauged models discussed by Antoniadis and
  Obers~\cite{Antoniadis:1994jx}.
 In the general case there is an
  additional term in the metric and a
  non-trivial three-form flux, both proportional to the difference
  $\mu-\mubb$. The Ricci scalar
\begin{equation}
  R\ =\ -\frac{4F_-(u)}{x^2\Delta(u)}\ \ ,
\end{equation}
clearly exhibits the singular nature of the background. Upon performing a T-duality transformation
along the $\f$ direction we obtain a plane wave background which
does not depend on $\mu$ and $\mubb$. The explicit form of the dual
background is
\begin{equation}
  ds^2
  \ =\ \frac{\Delta(u)}{F_-(u)} \left [ dx^2 + x^2 d \theta^2\right ]
       - 2dudv \ \ ,
\end{equation}
\begin{equation}
  B_{\theta x }\ =\ \frac{2 l_2(u) x }{F_-(u)}\ ,
  \hspace{1.4cm}
  \Phi\ =\ - \frac{1}{2} \ln\bigl[F_-(u)\bigr] \ \ ,
\end{equation}
where $\theta$ is the T-dual coordinate. The parameters $\m$ and
$\bar \m$ were removed by the coordinate transformation $v
\rightarrow v - 2\Lambda(\m - \bar \m) \theta$.

\section{\label{sc:RepTh}Coset characters and representation theory}

  The vertex operators of a coset conformal field theory $G/H$
  transform in irreducible representations of the coset chiral
  algebra. The goal of this section is to derive the characters of
  these representations in order to provide a precise
  description of the $\sigma$-model spectrum. Due to the
  non-compactness and non-semi-simplicity of the numerator group the
  standard methods of determining the branching functions fail. The
  reason for this failure may be attributed to unavoidable divergences
  which arise in the product of characters belonging to infinite
  dimensional representations of the horizontal subalgebra. In this
  section we propose a method to circumvent this problem. Our
  approach makes use of both character techniques and the knowledge
  of tensor products of the horizontal subalgebra. Since our method
  only rests on the absence of singular vectors on higher energy
  levels it should be applicable to general non-compact coset
  theories. We briefly comment on subtleties which arise in connection
  with the decomposition of tensor products involving spectral flow
  representations and the potential occurrence of representations that
  are not fully decomposable.

\subsection{\label{sc:scAnalysis}Semi-classical analysis}

  We begin this section with a discussion of the
  semi-classical approximation to string propagation on group
  manifolds and their cosets. This allows us  to
  introduce the unitary representations of the Heisenberg algebra
  $H_4$ and to illustrate in a simple context the relation between
  coset characters and branching functions.

  When all the length scales in a problem are large compared to the
  string scale, there is no significant difference between the
  behaviour of a string and the behaviour of a point particle. In this
  semi-classical approximation the Hilbert space of states for a
  string moving on a non-compact group manifold $G$ coincides with the
  space of functions $\mc{F}(G)$ which are $\delta$-function
  normalisable. This space admits a left-right-regular action of $G$
  and decomposes as
\begin{equation}
  \label{eq:HFunc}
  \mc{F}(G)\ =\ \int\!\!d\mu\,V_\mu\otimes V_\mu^\ast\ \ ,
\end{equation}
  where the direct integral runs over a certain set of unitary
  irreducible representations $\m$ of $G$. In the case of the
  Heisenberg group, $\mc{F}(H_4)$ can be written as a direct integral
  over three classes of representations~\cite{VilenkinBook}.
  There are two families of so-called
  discrete series representations $(\pm,p,j)$ (with $p > 0$ and
  $j\in\Real$) and one family of continuous series representations
  $(0,s,j)$ (with  $s\geq0$ and $j$ defined modulo $1$). We
    will use the same symbols later when we talk about the associated
    Lie algebra.

All these representations are infinite dimensional,
  thereby reflecting the non-compactness of the group manifold
  $H_4$. The $J$ and $K$ eigenvalues of the states in a given representation
  $\m$ are encoded by the characters
\begin{equation}
  \rho_\mu(z,w)
  \ =\ \tr_\mu\Bigr[z^{-iJ}w^{-iK}\Bigr]\ \ ,
\end{equation}
  whose explicit expressions are
\begin{equation}
  \label{hcharacters}
    \rho_{(+|p,j)}(z,w)
     =\ \frac{z^j\,w^p}{1-z} \ ,
     \hspace{0.6cm}
    \rho_{(-|p,j)}(z,w)
     =\ \frac{z^j\, w^{-p}}{1-z^{-1}} \ ,
     \hspace{0.6cm}
    \rho_{(0|s,j)}(z,w)
     =\ \sum_{n\in\Integer}z^{n+j}\ \ .
\end{equation}
  The characters should be thought of as formal sums counting the
  multiplicity of states with given quantum numbers.

  In a similar way the Hilbert space of a coset theory is given by
  $\mc{F}(G/H)$. This space coincides with the subspace of $\mc{F}(G)$
  consisting of the $H$-invariant functions
\be
  \Inv_H\mc{F}(G)\ :=\
  \bigl\{ f \in \mc{F}(G) \, \bigl| \, f(g)=f\bigl(\epsilon(h)g\bar \epsilon(h^{-1})\bigr)
  \, , \forall g \in G \, , \forall h \in H \bigr\} \ \ ,
\ee
  where
  $\epsilon$ and $\bar \epsilon$ denote the two embeddings of $H$ in
  $G$ used to define the coset. Since $\mc{F}(G)$ can be decomposed
  according to Eq.~\eqref{eq:HFunc}, we can obtain an explicit
  description of $\mc{F}(G/H)$ by first restricting all
  $G$-representations $V_\mu$ to $H$-representations
  $V_\mu\bigr|_H=\oplus_a{b_\mu}^aV_a$ and then taking the
  $H$-invariant part by coupling the tensor product of left and right
  factors to the trivial representation.

  In the case of diagonal embeddings the branching coefficients are
  just the tensor product coefficients. In order to deduce the
  branching functions for the diagonal cosets of the $H_4$ WZW model
  we will thus need the following tensor products of representations
  of the Heisenberg group~\cite{VilenkinBook}\footnote{As a function
    the product is obviously well-defined but not a priori as a
    generating function for the states in the tensor product
    representation.}
\begin{equation}
  \label{eq:TensorProducts}
  \begin{split}
    (\pm|p_1,j_1)\otimes(\pm|p_2,j_2)
    &\ =\ \bigoplus_{n=0}^\infty\,(\pm|p_1{+}p_2,j_1{+}j_2{\pm}n)\\[2mm]
    (\pm|p_1,j_1)\otimes(\mp|p_2,j_2)
    &\ =\ \begin{cases}  \bigoplus_{n=0}^\infty\,
    \bigl(\tau\bigr| |p_1-p_2|,j_1{+}j_2{-}\tau n\bigr)\, ,\ \tau=\pm\sign(p_1{-}p_2)&,\,p_1\neq p_2\\[2mm]
    \int_0^\infty\!ds\,s\,(0|s,j_1{+}j_2)&,\,p_1=p_2\end{cases}\\[2mm]
    (\pm|p_1,j_1)\otimes(0|\s,j_2)
    &\ =\ \bigoplus_{n\in\Integer}\,(\pm|p_1,j_1{+}j_2{+}n) \\[2mm]
(0|s_1,j_1)\otimes(0|s_2,j_2)
    &\ =\ \int_0^{2 \pi} \frac{d \psi}{2 \pi} \ (0|s(\psi),j_1{+}j_2) \ ,
    \hspace{0.6cm}  s^2(\psi) = s_1^2 + s_2^2 + 2
s_1 s_2 \cos \psi \ \ .\raisetag{66pt}
  \end{split}
\end{equation}
  Writing these tensor products in terms of characters one can derive
  some formal rules to interpret the following a priori ill-defined
  products
\begin{equation}
  \label{frules}
  \begin{split}
    \rho_{(+|p,j_1)} \, \rho_{(-|p,j_2)} &\ =\ %
 \frac{1}{1-z}\frac{1}{1-z^{-1}} \ := \ \int_0^\infty\!ds\,s\,
\sum_{n\in\Integer} \ z^{n+j_1+j_2} \ , \\[2mm]
\rho_{(0|s_1,j_1)} \, \rho_{(0|s_2,j_2)} &\ =\ \sum_{n, m
\in\Integer} \ z^{n+m+j_1+j_2}\ := \  \int_0^{2 \pi} \frac{d \psi}{2
\pi} \ \ \sum_{n\in\Integer} \ z^{n+j_1+j_2} \ \ .
  \end{split}
\end{equation}
  These rules will be a valuable aid below when it comes to
  decomposing certain products of affine characters.

\subsection{\label{sc:art} Affine representation theory}

  The symmetry algebra of the WZW model based on the Heisenberg group
  is generated by an affine $H_4$ algebra. In this section we define
  the $H_4$ algebra giving the commutation relations for the modes of
  the affine currents. We then discuss two classes of representations,
  the standard representations and the spectral flow representations
  \cite{Maldacena:2000hw}. The difference between the two is that the
  spectrum of the Virasoro generator $L_0$ is not bounded from below
  in the spectral flow representations.

\subsubsection{Standard representations}

The affine $\affH$ algebra is defined by the commutation
relations \be
  [P_n^+,P_m^-] \ =\ 2n\,\delta_{n+m,0}-2iK_{n+m} \ , \hspace{0.2cm}
  [J_n,P_m^\pm] \ =\ \mp i P_{n+m}^\pm \ , \hspace{0.2cm}
  [J_n,K_m] \ =\ n\,\delta_{n+m,0} \ ,
\ee with $n, m \in \mathbb{Z}$.
  The simplest class of irreducible representations of $\affH$ are the
  highest-weight representations, generated by acting with all the
  negative modes of the currents on an irreducible unitary
  representation $\mu$ of the horizontal subalgebra. Generalising the
  definition given for the horizontal subalgebra, we introduce the
  following characters
\begin{equation}
  \label{eq:CharDef}
  \chi_\mu^\affH(q,z,w)\ =\ \tr_\mu\Bigr[q^{L_0-\frac{c}{24}} \, z^{-iJ_0} \, w^{-iK_0}\Bigr]\
  \ .
\end{equation}
  Here $J_0$ and $K_0$ are the  zero modes of the corresponding affine
  currents and $L_0$ is the zero mode of the energy-momentum tensor
\be
  T = \frac{1}{2}\left ( P_1^2 + P_2^2 + J K + K^2 \right ) \ .
\label{emt}
\ee
  Due to the absence of singular vectors explicit expressions for the
  characters are easily computed
\begin{equation}
  \begin{split}
    \label{eq:AffineCharactersOne}
    \chi_{(+|p,j)}^\affH(q,z,w)
    &\ =\ \frac{q^{h_{(+,p,j)}-\frac{1}{12}}\,z^j\,w^p}{(1-z)\eta(q)^2
    \prod_{n=1}^\infty(1-zq^n)(1-z^{-1}q^n)}
     \hspace{10mm}\bigl(\,|q|<|z|<1\,\bigr)\\[2mm]
    \chi_{(-|p,j)}^\affH(q,z,w)
    &\ =\ \frac{q^{h_{(-,p,j)}-\frac{1}{12}}\,z^j\,w^{-p}}{(1-z^{-1})\eta(q)^2
    \prod_{n=1}^\infty(1-zq^n)(1-z^{-1}q^n)}
     \hspace{4mm}\bigl(\,|q|^{-1}>|z|>1\,\bigr)\\[2mm]
    \chi_{(0|s,j)}^\affH(q,z,w)
    &\ =\ \frac{q^{h_{(0,s,j)}-\frac{1}{12}}\,\sum_{n\in\Integer}z^{n+j}}{\eta(q)^2
    \prod_{n=1}^\infty(1-zq^n)(1-z^{-1}q^n)}
     \ =\ \frac{q^{h_{(0,s,j)}}}{\eta(q)^4}\sum_{n\in\Integer}z^{n+j}\ \ .
  \end{split}
\end{equation}
  The conformal weights of the ground states of these representations
  coincide with the eigenvalues of the modified Casimir operator
  associated with the energy-momentum tensor in Eq.~\eqref{emt}. They
  are given by
\begin{align}
  \label{eq:ConformalDimensions}
  h_{(\pm|p,j)}&\ =\ \frac{p}{2}(1-p)\mp pj\quad,&
  h_{(0|s,j)}&\ =\ \frac{s^2}{2}\ \ .
\end{align}
For future reference we also show how these expressions are modified
when the invariant metric on $H_4$ is not the standard one but has
the general form \eqref{eq:metric-g}. The conformal dimensions then
become
\begin{align}
 \label{eq:ConformalDimensions-d}
  h_{(\pm|p,j)}&\ =\ \frac{1}{\Lambda}\left(\frac{p}{2} \mp p j \right)+  \lambda
  p^2- \frac{p^2}{2 \Lambda^2},&
  h_{(0|s,j)}&\ =\ \frac{s^2}{2 \Lambda}\ \ .
\end{align}
As it is the case for the $SL(2,\mathbb{R})$ WZW model
  \cite{Maldacena:2000hw}, only a subset of the highest-weight
  representations is part of the spectrum of the theory. For $\affH$
  the allowed highest-weight representations are $(0|s,j)$ and
  $(\pm|p,j)$ with $p \in (0,1)$~\cite{Kiritsis:2002kz,
    D'Appollonio:2003dr}. In the following we will call them standard
  representations. States with $p \ge 1$ belong to a different class
  of representations called spectral flow
  representations~\cite{Maldacena:2000hw} in which $L_0$ is not bounded
  from below.

\subsubsection{Spectral flow representations}

  The name of this class of representations has its origin in the
  observation that the $\affH$ current algebra admits a family of
  spectral flow automorphisms $\Sigma_\omega$, $\w \in \mathbb{Z}$,
  which act on the modes as
\begin{align}
  \label{eq:SF}
  \Sigma_\omega(P_n^\pm)&\ =\ P_{n\mp \omega}^\pm\ ,&
  \Sigma_\omega(J_n)&\ =\ J_n\ ,&
  \Sigma_\omega(K_n)&\ =\ K_n-i\omega\delta_{n0}\ \ .
\end{align}
  From this definition one also readily derives the action
\begin{equation}
  \label{eq:SFVir}
  \Sigma_\omega(L_n)\ =\ L_n-i\omega J_n \ ,
\end{equation}
  on the Virasoro modes. Given a representation $\mu$ implemented on
  the space $\mc{H}_\mu$ via the map
  $\rho_\mu:\affH\to\End(\mc{H}_\mu)$, we can define a new
  representation $\mu_\omega$ which acts on the same space via the map
  $\rho_{\mu_\omega}=\rho_\mu\circ\Sigma_{-\omega}$. In view of its
  construction $\mu_\omega$ is called a spectral flow
  representation. Spectral flow representations also exist for affine
  Lie algebras based on compact real forms of finite dimensional
  semi-simple Lie algebras but in this case $L_0$ is still bounded
  from below and it can be shown that they are equivalent to ordinary
  highest-weight representations. This, however, is not the case for
  non-compact affine algebras and in particular for $\affH$. The
  inclusion of the spectral flow representations in the spectrum
  allows to extend the range of the label $p$ from the unit interval
  to the whole positive real axis.

  Using the equations \eqref{eq:SF} and \eqref{eq:SFVir} one can
  easily relate the character of $\mu_\omega$ to the character of the
  underlying standard representation $\mu$. Indeed, a simple algebraic
  manipulation within the trace yields
\begin{equation}
  \label{eq:SFChar}
  \begin{split}
    \chi_{\mu_\omega}^\affH(q,z,w)
    \ =\ w^{\omega}\,\chi_\mu^\affH(q,q^{-\omega}z,w)\ \ .
  \end{split}
\end{equation}
  In order to simplify the notation we will henceforth identify the
  label $\mu_{\omega=0}$ with $\mu$ whenever there is no chance of
  confusion.

\subsection{\label{sc:RepStructure} Tensor product decompositions
for the diagonal coset}

  In this section we analyse the decomposition of the $\affH \times \affH$ representations with
  respect to the diagonal subalgebras $\e\bigl(\affH\bigr)$ which are relevant to the
  curved backgrounds discussed earlier in this paper. We explain why
  the standard character decompositions fail and provide a method
  which allows to circumvent these problems by using a mixture of
  character techniques and analytical input from tensor product
  decompositions of the horizontal subalgebra.

\subsubsection{General strategy}

  The affine standard representations
  relevant for the $H_4$ WZW model are all induced from
  infinite dimensional unitary representations of the horizontal subalgebra.
  The modes, however, which are used to generate the remaining states transform
  in the finite dimensional adjoint representation which is {\em non-unitary},
  a common feature of all WZW models based on non-compact groups.
  Yet, in the present case there is an additional complication because the
  adjoint representation is {\em reducible but not fully decomposable}, reflecting
  the non-semi-simple nature of the Lie algebra $H_4$.

  In this section we describe a method to derive the decomposition of
  the tensor products of standard affine representations. Our general
  strategy is to decompose the affine representations into
  representations of the horizontal subalgebra on each energy level
  first. Then we use the known tensor products for the horizontal
  subalgebra in order to determine the tensor product energy level by
  energy level. Finally we reorganise the result and express it in
  terms of affine characters again. This last step is in fact greatly
  simplified by the absence of singular vectors in the affine modules
  that are relevant here as we will explain below.

  The main advantage of the method just described is that it allows to
  combine character techniques with the analytic knowledge about the
  tensor products of the horizontal subalgebra displayed in
  Eq.~\eqref{eq:TensorProducts}. This is very convenient for
  non-compact groups since the unitary representations are infinite
  dimensional. Unfortunately this method cannot be applied directly to
  the spectral flow representations, as discussed in more detail in
  section~\ref{sc:CosetSF}. It also fails if a given tensor product
  turns out not to be fully decomposable.

  As already mentioned, the standard modules relevant for the
  WZW models are simply obtained by applying (properly symmetrised)
  combinations of negative modes to the ground states. Together with
  the absence of null vectors in the resulting Verma modules this
  allows us to represent the standard affine representations
  $\hat{\mu}$ in the form
\begin{equation}
  \label{eq:RepStructure}
  \hat{\mu}\bigr|_{H_4}\ =\ q^{h_\mu}\,\mu\otimes M(q)\ \ .
\end{equation}
  Here, $\mu$ is the underlying representation of the horizontal
  subalgebra and $M(q)$ denotes the universal enveloping algebra of
  the subalgebra generated by the negative modes of the
  $\affH$-currents. The variable $q$ keeps track of the energies of
  the states. Since all the modes of the affine currents transform in
  the adjoint representation we can write
\begin{equation}
  \label{eq:RepM}
  M(q)\ =\ 1+q\,\ad+q^2\bigl[\ad+(\ad\otimes\ad)_\sym\bigr]+q^3
  \bigl[\ad+\ad\otimes\ad+(\ad\otimes\ad\otimes\ad)_\sym\bigr]+\cdots\ \ .
\end{equation}
  As discussed in appendix~\ref{ap:AdTensor}, the tensor products of
  the adjoint representation contain indecomposable but reducible
  representations. However it is easy to see that the tensor product
  $\ad \otimes (\pm|p,j)$ is fully reducible. In fact indecomposable
  representations can only appear when the eigenvalue of $K_0$
  vanishes in the tensor product. The three examples of not fully
  decomposable representations relevant for us
  are the tensor products $\ad^{\otimes \, n}$, $(0|s_1, j_1) \otimes
  (0|s_2, j_2)$ and $(+|p, j_1) \otimes (-|p, j_2)$. These cases are
  also analysed in more detail in appendix~\ref{ap:AdTensor}.

  From the previous paragraph we conclude that the product in
  \eqref{eq:RepStructure} is fully reducible when $\mu = (\pm|p,j)$
  and we obtain
\begin{equation}
  \label{eq:SimpleTensor}
  (\pm|p,j)\otimes M(q)
  \ =\ \bigoplus_{n\in\Integer}N_{[M,n]}^\pm(q)\ (\pm|p,j+n)\ \ .
\end{equation}
  We can derive the explicit form of the multiplicity functions
  $N_{[M,n]}^\pm(q)$ by writing the previous equation in terms of
  characters. The character of $M(q)$ is given by
\begin{equation}
  \chi_M(q,z)
  \ =\ \prod_{n=1}^\infty\Bigl[(1-q^n)^2(1-zq^n)(1-z^{-1}q^n)\Bigr]^{-1}\ \ .
\end{equation}
For $|q|<|z|<1$ a more convenient form is  \cite{Pakman:2003kh,
D'Appollonio:2007bn}
\begin{equation}
   \chi_M(q,z)
   \ =\ \sum_{n\in\Integer}z^n\sum_{m=1}^\infty(-1)^{m+1}
    \frac{q^{\frac{m}{2}(m+2n-1)+\frac{1}{6}}(1-q^m)}{\eta(q)^4}\ \ .
\end{equation}
 Since this function is symmetric with respect to
  the replacement $z\mapsto1/z$  we can also write
\begin{equation}
  \label{eq:MChar}
  \chi_M(q,z)
  \ =\ \sum_{n\in\Integer}z^n\sum_{m=1}^\infty(-1)^{m+1}
  \frac{q^{\frac{m}{2}(m + 2|n|-1)+\frac{1}{6}}(1-q^m)}{\eta(q)^4}\
  \ .
\end{equation}
Substituting the previous expressions for $ \chi_M(q,z)$ in
Eq.~\eqref{eq:SimpleTensor} and comparing the coefficients of
identical powers of the variable $z$ on both sides of the equation
we obtain
\begin{equation}
  N_{[M,n]}^\pm(q)
  \ =\ \sum_{m=1}^\infty(-1)^{m+1}
  \frac{q^{\frac{m}{2}(m+2|n|-1)+\frac{1}{6}}(1-q^m)}{\eta(q)^4}\ \ .
\end{equation}
 Note that, as anticipated by our notation, the result does not
 depend on $j$.

When $\m = (0|s,j)$ we cannot follow the same approach because the
tensor products  $(0|s,j)\otimes\ad$ are reducible but not fully
decomposable (see appendix \ref{ap:AdTensor}). Similar problems can
  be expected for all non-compact groups and their non-abelian
  cosets. For instance in the case of $\widetilde{SL}(2,\Real)$
  the non-complete reducibility enters on sufficiently
  high energy levels in the discrete series of affine
  representations with half-integral spin.

We now apply the decomposition of the affine modules in
\eqref{eq:RepStructure} to the tensor product
  of two affine representations $ \hat{\mu}\otimes\hat{\nu}$, in order to
compute the branching functions $b_{[\mu,\nu,\sigma]}(q)$ in the
tensor product decomposition $ \hat{\mu}\otimes\hat{\nu}
  =\bigoplus_\sigma b_{[\mu,\nu,\sigma]}(q)\ \hat{\sigma}$.
We obtain
\begin{equation}
  \label{eq:affTP}
  \hat{\mu}\otimes\hat{\nu}\bigr|_{H_4}
  \ =\ q^{h_\mu+h_\nu}\ \mu\otimes\nu\otimes M(q)^2
  \ =\ \bigl(\,\mu\otimes\nu\otimes
  M(q)\,\bigr)^{\widehat{}}\bigr|_{H_4}\ \ ,
\end{equation}
where the hat over the
  tensor product representation on the right hand side indicates
  the affinisation of the $H_4$-representation $\mu\otimes\nu\otimes M(q)$.
  This affinisation should be understood as an induced $\affH$-module
  based on the given representation of the horizontal subalgebra.
We are assuming that the affine
  representations on both sides of the previous equation
  contain the {\em same} factor $M(q)$ coming from the higher modes.
  This is true because the Verma modules are irreducible.
To evaluate $\mu\otimes\nu\otimes M(q)$ we first perform
  the tensor product $\mu\otimes\nu=\oplus_\sigma N_{\mu\nu}^\sigma\,\sigma$
  using eqs.\ \eqref{eq:TensorProducts} and then calculate energy level by energy level
  the tensor product of infinite dimensional representations with finite dimensional ones
  using character techniques.
  The final result is schematically given by
\begin{equation}
  \mu\otimes\nu\otimes M(q)
  \ =\ \bigoplus_{\sigma,\rho}N_{\mu\nu}^\sigma N_{\sigma M}^\rho(q)\,\rho
  \ =\ \bigoplus_\rho b_{[\mu,\nu,\rho]}(q)\,\rho\ \ .
\end{equation}

\subsubsection{\label{sc:CosetPPP}Decomposition of the tensor products of standard representations}

We now apply the procedure just outlined to the decomposition of the
  $\affH\times\affH$ representations with respect to the diagonal
  $\affH$ subalgebras discussed in section~\ref{sc:Data}.
  With no loss of generality we may assume
  that the embedding is of the form \eqref{eq:Embedding} with
  $\eta_i=1$. All other choices can be reduced to this one by
  a suitable automorphism.

Before giving the general result, we derive in some detail the
characters of the representations of the coset chiral algebra that
appear in $(+|p_1,j_1)\otimes(+|p_2,j_2)$. We first compute the
following tensor products
\ba
    & & (+|p_1,j_1)\otimes(+|p_2,j_2)\otimes M(q)
    \ =\ \bigoplus_{n=0}^\infty(+|\nu_1^2p_1+\nu_2^2p_2,j_1+j_2+n)\otimes M(q) \nb  \\
    \ &=&\ \bigoplus_{n=0}^\infty\bigoplus_{l\in\Integer}(+,\nu_1^2p_1+\nu_2^2p_2,j_1+j_2+n+l)
          \sum_{m=1}(-1)^{m+1}\frac{q^{\frac{m}{2}(m-2l-1)+\frac{1}{6}}(1-q^m)}{\eta(q)^4}  \nb \\
\ &=&\ \bigoplus_{l\in\Integer}(+|\nu_1^2p_1+\nu_2^2p_2,j_1+j_2+l)
          \sum_{m=1}(-1)^{m+1}\frac{q^{\frac{m}{2}(m-2l-1)+\frac{1}{6}}}{\eta(q)^4}\ \ .
\ea
  Then we include all the factors $q^{h_\mu}$ and $q^{-
\frac{c}{24}}$ required by the definitions  \eqref{eq:CharDef} and
\eqref{eq:RepStructure}. The final result for the coset character is
\begin{equation}
  \label{eq:DecoPPP}
  \chi_{[(+|p_1,j_1),(+|p_2,j_2);(+|p,j_1+j_2+n)]}^\HHH(q)
  \ =\ \frac{q^{h_{(+|p_1,j_1)}+h_{(+|p_2,j_2)}-h_{(+|p,j_1+j_2+n)}}}
  {\eta(q)^4} \ \sum_{m=0}^\infty(-1)^m q^{\frac{m}{2}(m+2n+1)}\ \ ,
\end{equation}
with $p=\nu_1^2p_1+\nu_2^2p_2$. From the previous expression we can
read off the conformal dimension of the coset primary fields
\begin{equation}
  h_{[(+|p_1,j_1),(+|p_2,j_2);(+|p,j_1+j_2+n)]}^\HHH
  \ =\ \begin{cases}
         h_{(+|p_1,j_1)}+h_{(+|p_2,j_2)}-h_{(+|p,j_1+j_2+n)}&, \text{ for }n\geq0\\[2mm]
         h_{(+|p_1,j_1)}+h_{(+|p_2,j_2)}-h_{(+|p,j_1+j_2+n)}-n&, \text{ for }n\leq0\ \ .

       \end{cases}
\end{equation}
\smallskip
All other cases of the form $\mu_1 \otimes \mu_2$ with $\mu_1 =
(\pm|p_1,j_1)$ and $\mu_2 = (\pm|p_2,j_2)$ can be treated in exactly
the same way. The result can be written in the following compact
form
\begin{equation}
  \chi_{[\mu_1,\mu_2;\mu_{1 2}(n)]}^\HHH(q)
  \ =\ \frac{q^{h_{\mu_1}+h_{\mu_2}-h_{\mu_{1  2}(n)}}}
  {\eta(q)^4} \ \sum_{m=0}^\infty(-1)^m q^{\frac{m}{2}(m+2n+1)} \ ,
\label{eq:deco-gen}
\end{equation}
\begin{equation}
  h_{[\m_1, \m_2, \m_{1  2}(n)]}^\HHH
  \ =\ \begin{cases}
         h_{\m_1}+h_{\m_2}-h_{\m_{1  2}(n)}&, \text{ for }n\geq0\\[2mm]
         h_{\m_1}+h_{\m_2}-h_{\m_{1  2}(n)} - n &, \text{ for }n\leq0\ \ .
       \end{cases}
\end{equation}
The label $\mu_{1  2}(n)$ is specified by the following rules that
simply reflect the tensor products of the horizontal algebra
\begin{align}
\mu_1 \ \ \ \ &  & \mu_2 \ \ \ \ & & p > 0 \ \ \ & &\mu_{1  2}(n)   \ \ \ \ \ \ \nb \\
(+|p_1,j_1) & & (+|p_2,j_2) & & \nu_1^2p_1+\nu_2^2p_2 & &
(+|p,j_1+j_2+n) \nb \\
(-|p_1,j_1) & & (-|p_2,j_2) & & \nu_1^2p_1+\nu_2^2p_2 & &
(-|p,j_1+j_2-n) \nb \\
(+|p_1,j_1) & & (-|p_2,j_2) & & \nu_1^2p_1-\nu_2^2p_2 & &
(+|p,j_1+j_2-n) \nb \\
(+|p_1,j_1) & & (-|p_2,j_2) & & \nu_2^2p_2-\nu_1^2p_1 & &
(-|p,j_1+j_2+n) \nb
\end{align}
  In the previous and in the following formulas the conformal
  dimension of the representations of the embedded algebra are given
  by \eqref{eq:ConformalDimensions-d}, since one should use the
  induced metric on $\e\bigl(H_4\bigr)$.

  The only cases that require a different approach are the product
  $(+|p_1,j_1)\otimes(-|p_2,j_2)$ with $\nu_1^2p_1 - \nu_2^2p_2 = 0$
  and the product $(0|s_1,j_1)\otimes(0|s_2,j_2)$. In both cases the
  full reducibility of the induced module $\bigl(\m \otimes \n\otimes
  M(q)\bigr)^{\widehat{}}$ is not guaranteed since the tensor products
  $(0|s,j)\otimes\ad$ are reducible but not fully decomposable. A
  priori one cannot exclude that indecomposable affine representations
  could play a role in the construction of the diagonal $H_4$
  cosets. This could be a rather common feature of non-compact cosets
  involving a non-abelian denominator and a closer investigation of
  this phenomenon and of its possible connections with logarithmic
  conformal field theories is left for further work. Let us mention
  that at least in the case of $H_4$ the indecomposable
  representations, if present, would not be part of the physical
  string spectrum since they will be removed by the Virasoro
  constraints.

  Although we cannot provide a rigorous discussion of the
  decomposition of $(+|p_1,j_1)\otimes(-|p_2,j_2)$ with $\nu_1^2p_1 -
  \nu_2^2p_2 = 0$ and $(0|s_1,j_1)\otimes(0|s_2,j_2)$, we can derive a
  simple and plausible expression for the coset characters {\it
    assuming} the full reducibility of the tensor product of the
  affine representations and using the formal rules in
  Eq.~\eqref{frules}. In the first case the full reducibility
  translates into the following character identity
\be
 \chi_{(+|p_1,j_1)}^\affH(q,z)  \chi_{(-|p_2,j_2)}^\affH(q,z) =
 \int_0^\infty ds \, s \ \chi_{[(+|p_1,j_1),(+|p_2,j_2);(0|s,j_1+j_2)]}^\HHH(q)
\ \chi_{(0|s,j_1+j_2)}^\affH(q,z) \ , \ee
  and after simplifying the
common factors on both sides we obtain \be
\chi_{[(+|p_1,j_1),(+|p_2,j_2);(0|s,j_1+j_2)]}^\HHH(q) \ =\
\frac{q^{h_{(+|p_1,j_1)}+h_{(-|p_2,j_2)}-h_{(0|s,j_1+j_2)}}}{\eta(q)^4}\
\  ,\quad \nu_1^2p_1 - \nu_2^2p_2 = 0 \ . \label{eq:decopm}
\end{equation}
As for the second case, we start from \be
 \chi_{(0|s_1,j_1)}^\affH(q,z)  \chi_{(0|s_2,j_2)}^\affH(q,z) =
 \int_0^{2 \pi} \frac{d \psi}{2 \pi}  \ \chi_{[(0|s_1,j_1),(0|s_2,j_2);(0|s(\psi),j_1+j_2)]}^\HHH(q)
\ \chi_{(0|s(\psi),j_1+j_2)}^\affH(q,z) \ , \label{eq:decozz} \ee
and using again Eq.~\eqref{frules} we obtain \be
\chi_{[(0|s_1,j_1),(0|s_2,j_2);(0|s(\psi),j_1+j_2)]}^\HHH(q) \ =\
\frac{q^{h_{(0|s_1,j_1)}+h_{(0|s_2,j_2)}-h_{(0|s(\psi),j_1+j_2)}}}{\eta(q)^4}\
\  ,
\end{equation}
where $s^2(\psi) = s_1^2 + s_2^2 + 2 s_1 s_2 \cos \psi$.

  The case $(\pm|p,j_1)\otimes(0|s,j_2)$ can again be discussed
  rigorously. However, the discussion in terms of character techniques
  is much simpler and leads directly to the result
\begin{equation}
  \chi_{[(\pm|p,j_1),(0|s,j_2);(\pm|p,j_1+j_2+n)]}(q)
  \ =\ \frac{q^{h_{(\pm|p,j_1)}+h_{(0|s,j_2)}-h_{(\pm|p,j_1+j_2+n)}}}{\eta(q)^4}\ \ .
\label{eq:decopz}
\end{equation}
  The conformal dimension is
\begin{equation}
  h_{[(\pm|p,j_1),(0|s,j_2);(\pm|p,j_1+j_2+n)]}^\HHH
  \ =\ h_{(\pm|p,j_1)}+h_{(0|s,j_2)}-h_{(\pm|p,j_1+j_2+n)}\ \ .
\end{equation}
  We stress again that the conformal dimension of the representations
  of the embedded algebra are computed with the induced metric and
  therefore are given by \eqref{eq:ConformalDimensions-d}.

\subsubsection{\label{sc:CosetSF}Remarks on the decomposition of tensor products with spectral flow representations}

  In the last part of this section we would like to comment on
  some aspects of the decomposition of tensor products involving spectral
  flow representations $\m_{\w_1} \otimes \nu_{\w_2}$. Our first
  observation is that the spectral flow
  automorphisms \eqref{eq:SF} and the embeddings  \eqref{eq:Embedding}
  satisfy the relation
\begin{equation}
  \label{eq:SFEmb}
   \epsilon\circ \bigl(\Sigma_{\eta_1\omega}\times\Sigma_{\eta_2\omega}\bigr)
  \ =\ \Sigma_\omega \circ\epsilon \ .
\end{equation}
Given a decomposition
\begin{equation}
  \mu \otimes \nu
  \ = \  \bigoplus_\lambda N_{\mu\nu}^\lambda\,\lambda \ , \ee
the previous relation implies \be
 \mu_{\eta_1\omega} \otimes \nu_{\eta_2\omega}
  \ = \  \bigoplus_\lambda N_{\mu\nu}^\lambda\,\lambda_{\omega} \  \
  , \hspace{1cm} \w \in \mathbb{Z} \ .
\end{equation}
The equivalence of coset characters resulting from
Eq.~\eqref{eq:SFEmb} is just a manifestation of what is known as
field identification in compact coset models. Indeed, it has been
known for a long time that field identifications are implemented by
the action of certain pairs of simple currents in the numerator and
the denominator affine algebra
\cite{Gepner:1989jq,Schellekens:1989uf}. Simple currents in turn can be
identified with spectral flow transformations and therefore
Eq.~\eqref{eq:SFEmb} precisely singles out the pairs of spectral
flows that induce field identifications. Since in contrast to the
compact case here we have to identify an infinite number of coset
representations, this leads to an infinite degeneracy in the coset
partition function which has to be removed by hand.

According to Eq.~\eqref{eq:SFEmb} and the corresponding field
identification, it is sufficient to consider only tensor products of
the form $\m \otimes \n_{\w}$, with $\m$ a standard representation
and $\n_{\w}$ a spectral flow representation. From this point of
view, the results obtained in the previous sections provide the
decompositions of the tensor products with $\w =0$ and, for the
discrete series, $\w = \pm 1$.
The analysis of the decomposition of the other tensor product $\m
\otimes \n_{\w}$ with $|\w| \ge 2$ is however a much more difficult
problem and it would be interesting to develop rigorous methods to
solve it. Character methods cannot be directly applied to this case.
The characters in Eq.~\eqref{eq:AffineCharactersOne} are in fact
formal power series that converge in different domains of the $z$
complex plane. For instance, using Eq.~\eqref{eq:SFChar}, we can see
that $\chi_{(+|p,j)_\omega}^\affH(q,z)$ converges in the annulus
$|q|^{\w+1} < |z| < |q|^\w$. As discussed in \cite{Lesage:2003kn}
the formal character and the analytic expression coincide only up to
contact terms that encode the unbounded part of the spectrum. At the
present stage it is
not obvious that one can find a consistent way of computing with
these formal series in order to extract the coset characters from
their product. This question provides an interesting direction for
future research.

\section{\label{sc:Spectrum}The spectrum of the diagonal cosets}

  In the final section of this paper we determine the operator content
  of the diagonal cosets that have been discussed in section
  \ref{sc:CosetGeometries}. If we combine these models with other CFTs
  such that the total central charge is the one required for a
  critical string theory background, the coset vertex operators
  correspond to closed string states propagating in the
  curved space-time described by the $\s$-model.

  The partition function of the WZW model based on $H_4 \times H_4$ is
  given by the charge conjugation modular invariant which couples
  every representation of the affine algebra with its conjugate
  representation. Hence any multiplet of primary fields is completely
  specified by fixing its transformation properties under the
  holomorphic affine current algebra. As we reviewed in section
  \ref{sc:RepTh} there are three types of representations of
  $\hat{H}_4$. For the derivation of the spectrum of the coset models
  it is convenient to divide the spectrum of $H_4 \times H_4$ into
  sectors labelled by the representations of the two $H_4$ factors. The
  spectrum then contains contributions from nine different sectors,
\begin{align} {\cal H}_{++} &= [(+| p_1, j_1),(+| p_2, j_2)]
& (0,0)<(p_1,p_2)<(1,1)
  & & (j_1, j_2) \in \mathbb{R}^2 \nb \\
{\cal H}_{+-} &= [(+| p_1, j_1),(-| p_2, j_2)] &  (0,0)<(p_1,p_2)<(1,1)  & & (j_1, j_2) \in \mathbb{R}^2 \nb \\
{\cal H}_{\pm 0} &= [(+| p_1, j_1),(\, 0| \, s, j_2)]  &  0 < p_1
<
1, \ \ \ s \ge 0  & & j_1 \in \mathbb{R}, \ \ 0 \le j_2 < 1 \nb \\
{\cal H}_{00} &= [(0| \, s_1, j_1),(0| \, s_2, j_2)] & s_1 \ge 0, \
\ s_2 \ge 0 & & (0,0) \le (j_1, j_2) < (1,1)
\end{align}
and similar definitions for the sectors ${\cal H}_{--}$, ${\cal
  H}_{-+}$ and ${\cal H}_{0 \pm}$. Moreover we have to take into
account the images of all these sectors under independent amounts of
spectral flow for the two $H_4$ factors.

  Our strategy for determining the spectrum of the coset theories
  discussed in this paper is as follows. For each of the sectors
  $\mc{H}_{\mu\nu}$ of $H_4 \times H_4$ we first calculate the
  modified tensor products $\mu\otimes_{\epsilon}\nu$ and
  $\mu\otimes_{\bar \epsilon}\nu$, which are defined using the
  diagonal embeddings $\epsilon$  and $\bar \epsilon$ instead of the
  standard coproduct. Since the two embeddings are in general
  different, this gives rise to different types of representations for
  the left and the right movers. In order to identify the states of
  the coset we then impose the constraint
\be
  \epsilon(X)+\bar{\epsilon}(X)=0 \ , \label{eq:cosetconstraint}
\ee
  for all generators $X\in\hat{H}_4$. This implies that in each sector
  we are only allowed to keep those contributions for which the labels
  of the left-moving and right-moving representations that result from
  the decomposition coincide. Another consequence of the constraint
  \eqref{eq:cosetconstraint} is that the operators in the spectrum of
  the coset models are completely identified by three labels, two for
  the representations of the original $H_4 \times H_4$ model and one
  for either of the representations of the embedded $H_4$. As we will see
  condition \eqref{eq:cosetconstraint} with $X=K_0$ will severely
  restrict the type of sectors that can contribute to the spectrum of
  the different cosets.

  Since the decomposition
  of the tensor product of spectral flow
  representations is still an open problem, the discussion in this section
  will be restricted to the standard representations. We also
  introduce the short hand notation
\be
  c = \cos \alpha \ , \hspace{1.4cm}  \bar c = \cos \bar
\alpha \ , \hspace{1.4cm}  s = \sin \alpha \ , \hspace{1.4cm}  \bar
s = \sin \bar \alpha \ \ , \label{eq:shorts}
\ee
  and set $\Lambda=1$ for notational convenience. It can easily be
  recovered by rescaling the parameters $\lambda$, $\mu$ and $\mubb$.

\subsection{The cosets of type $(--)$}

We now apply  the
  procedure outlined above to the models in the class $(--)$.
The first step is to identify in which of the sectors
$\mc{H}_{\m\n}$ it is possible to solve the constraint
\eqref{eq:cosetconstraint} with $X = K_0$. Let us consider for
instance the $\mc{H}_{++}$ sector. Using the
  embeddings $\e$ and $\bar \e$ in \eqref{eq:Embedding}, on the left
  we obtain representations of the form $(+|p,j)$ with $p=c^2p_1+s^2p_2$
  and on the right, due to the presence of the signs $\bar{\eta}_i=-1$, we
  obtain representations of the form  $(-|\bar{p},\bar{\jmath})$ with
  $\bar{p} = \bar{c}^2p_1+\bar{s}^2p_2$. Given the difference in
  sign, these representations can never coincide and we conclude
  that the sector $\mc{H}_{++}$ does not contribute  to the spectrum
  of the coset. In the same way one can exclude also the sectors
$\mc{H}_{--}$, $\mc{H}_{\pm0}$ and $\mc{H}_{0\pm}$.

The sectors that contribute to the spectrum are $\mc{H}_{+-}$,
$\mc{H}_{-+}$ and $\mc{H}_{00}$. The operator content of the coset
depends on the range of $\alpha$ and $\bar \alpha$ and we can
summarise the result of the analysis in the following schematic way
\begin{equation}
  \begin{split}
    \alpha<\balpha:\quad \ \ \ \ &
        \bigl[(\pm|p_1,j_1),(\mp|p_2,j_2);(\pm|p,j)\bigr]\\[2mm]
    \alpha=\balpha:\quad \ \ \ \ &
        \bigl[(\pm|p_1,j_1),(\mp|p_2,j_2);(0|s,j)\bigr]\\[2mm]
    \alpha>\balpha:\quad \ \ \ \ &
        \bigl[(\pm|p_1,j_1),(\mp|p_2,j_2);(\mp|p,j)\bigr]\\[2mm]
    \text{any }\alpha,\balpha:\quad \ \ \ \ &
      \bigl[(0|s_1,j_1),(0|s_2,j_2);(0|s,j)\bigr]\ \ .
  \end{split}
  \label{eq:schema--}
\end{equation}
Here the terms in the square brackets are the three labels used to
identify the coset characters. Note that there is a drastic change
in the type of coset representations that appear in the spectrum  as
we move across the line $\alpha=\balpha$ which corresponds to the
singular geometries discussed in section \ref{sc:CosetGeometries}.

  We now solve the constraints $\epsilon(X)+\bar{\epsilon}(X)=0$
  associated with the other generators using the coset decompositions
  derived in section \ref{sc:CosetPPP}. In this way we can
  determine the relations among the parameters of the various
  representations in \eqref{eq:schema--}. We perform this analysis
  separately for the $\mc{H}_{\pm\mp}$ and the $\mc{H}_{00}$
  sectors.
\bigskip

  \noindent{\em The sector $\mc{H}_{\pm\mp}$:}
  Let us begin with the case
  $\alpha\neq\balpha$. We consider first the labels $p$ and $\bar p$.
  On the left we have  $p=\tau(c^2p_1-s^2p_2)$ and on the right
  $\bar{p}=-\tau(\bar{c}^2p_1-\bar{s}^2p_2)$ where $\tau=+1$
  for $\alpha<\balpha$ and $\tau=-1$ for $\alpha>\balpha$. This
  corresponds to a coset representation of the form
  $\bigl[(\pm|p_1,j_1),(\mp|p_2,j_2);(\pm\tau|p,j)\bigr]$. The
  equation $p=\bar{p}$ which follows from the coset constraints
  can be solved for $p_2$, giving
\begin{equation}
  p_2\ =\ \frac{c^2+\bar{c}^2}{s^2+\bar{s}^2}\,p_1
     \ =:\ rp_1\ \ ,
\label{pmm}
\end{equation}
and therefore
\begin{equation}
  p\ =\ \tau(c^2p_1-s^2p_2)
   \ =\ \frac{c^2\bar{s}^2-s^2\bar{c}^2}{s^2+\bar{s^2}}\,\tau p_1
   \ =\ -\frac{\sin(\alpha-\bar{\alpha})\sin(\alpha+\bar{\alpha})}{\sin^2\alpha+\sin^2\balpha}\,\tau p_1\ \ .
\end{equation}
  This is always positive, as required by the consistency of our
decomposition. Next we can determine the allowed values of $j$. To do
so we have to solve the equation
\begin{equation}
  j\ :=\ j_1+j_2\pm\mu p_1\mp(\lambda-\mu)p_2\mp\tau n
   \ =\ -(j_1+j_2)\mp\bar{\mu}p_1\pm(\lambda-\bar{\mu})p_2\mp\tau\bar{n}\ \ .
\end{equation}
  The integers $n,\bar{n}$ arise from the tensor product
  decomposition in Eq.~\eqref{eq:deco-gen}. Given concrete values for $p_1$, $j_1$, $n$ and
  $\bar{n}$, this equation can always be solved for $j_2$, resulting
  in
\begin{equation}
  \begin{split}
    j_2&\ =\ -j_1\pm\tau(n-\bar{n})/2\mp\bigl[(\mu+\bar{\mu})(1+r)-2\lambda r\bigr]\,p_1/2\\[2mm]
    j&\ =\ \mp\tau(n+\bar{n})/2\pm(\mu-\mubb)(1+r)p_1/2\ \ .
  \end{split}
\end{equation}
  Therefore when $\alpha \ne \bar \alpha$ all the parameters in \eqref{eq:schema--}
  can be expressed in terms of
  the data $(p_1,j_1,n,\bar{n})$.
\smallskip

  When $\alpha=\balpha$, from \eqref{pmm} it follows that $p_2 = \cot^2 \alpha \, \, p_1$ and
  that the sector $\mc{H}_{\pm\mp}$ decomposes into
  continuous representations. We then find the conditions
\ba j\ &=& \ j_1+j_2\pm\mu p_1\mp(\lambda-\mu)p_2  \ \ \ \ \ \mod\Integer \  , \nb \\
\bar \jmath\ &=& \ -(j_1+j_2)\mp\bar{\mu}p_1\pm(\lambda-\bar{\mu})p_2
\ \ \ \ \ \mod\Integer \ , \label{jmm} \ea with $j = \bar \jmath$
or $j = 1 - \bar \jmath$. In the first case the equation can be
solved by \ba j_2\ &=&\ -j_1 + n/2 \mp
\bigl[(\mu+\bar{\mu})(1+r)-2\lambda r\bigr]\,p_1/2 \ , \nb \\
j &=& n/2 \pm (\mu-\bar{\mu})(1+r)\,p_1/2  \ \ \ \ \ \mod\Integer \
, \ea with $n \in \mathbb{Z}$. In the second case if $\m = \bar \m$
there is no restriction on $j_1$ and $j_2$ and $j$ is given by
\eqref{jmm}, while if  $\m \ne \bar \m$ there are solutions only
when $(\m - \bar \m)(1+r) p_1 \in \mathbb{Z}$. Finally there is no
  restriction on the allowed range of $s$ as one can see from
  Eq.~\eqref{eq:decopm}.
\bigskip

  \noindent{\em The sector $\mc{H}_{00}$:}
  In this sector we have the constraint
\begin{equation}
  \label{eq:jconstraint}
  j\ =\ j_1+j_2\ \equiv\ -(j_1+j_2)\mod\Integer\ \ .
\end{equation}
  This equation can be solved by $j_2=\Upsilon-j_1$ with
  $\Upsilon\in\{1/2,1\}$, implying $j=\Upsilon$. The range of $s$
  follows from the expression \eqref{eq:TensorProducts} for the tensor
  product of the continuous representations, taking into account the
  action of the embeddings. The result is $ s_{\text{min}} \, \leq s
  \leq \, s_{\text{max}}$ where the upper and lower bounds are given by
\be
  \label{eq:Borders}
  s_{\text{min}}\ =\ \max\left (|c s_1 - s s_2|, |\bar c s_1 - \bar s s_2| \right
) \ , \hspace{1cm}
 s_{\text{max}} = \min\left ( c s_1 + s s_2, \bar c s_1 + \bar s s_2 \right
 ) \ \ .
\ee
  This concludes our discussion of the spectrum of cosets of type
  $(--)$.

\subsection{The cosets of type $(+-)$}

  The whole discussion for this class of models mimics the one in the
  previous subsection. In particular, we again have to distinguish
  three cases. Depending on the relative value of the parameters
  $\alpha$ and $\balpha$ we find the sectors
\begin{equation}
  \begin{split}
    \alpha<\balpha:\quad&
        \bigl[(\pm|p_1,j_1),(\mp|p_2,j_2);(\pm|p,j)\bigr]\\[2mm]
    \alpha=\balpha:\quad&
        \bigl[(\pm|p_1,j_1),(0|s,j_2);(\pm|s,j)\bigr]\\[2mm]
    \alpha>\balpha:\quad&
        \bigl[(\pm|p_1,j_1),(\pm|p_2,j_2);(\pm|p,j)\bigr]\\[2mm]
    \text{any }\alpha,\balpha:\quad&
      \bigl[(0|s_1,j_1),(0|s_2,j_2);(0|s,j)\bigr]\ \ .
\label{eq:schemapm}
  \end{split}
\end{equation}
Note that in this case different sectors $\mc{H}_{\m\n}$ of the $H_4
\times H_4$ model contribute for different values of the parameters
$\alpha$ and $\bar \alpha$. More precisely, besides the
$\mc{H}_{00}$ sector that is always in the spectrum,  we have the
$\mc{H}_{\pm\mp}$ sectors when $\alpha<\balpha$,  the
$\mc{H}_{\pm\pm}$ sectors when $\alpha>\balpha$ and finally  the
$\mc{H}_{\pm 0}$ sectors when $\alpha = \balpha$.
  In order to complete the description we have to
  impose the constraints and derive the
  relations between the different parameters in \eqref{eq:schemapm}.%
\bigskip

  \noindent{\em The sector $\mc{H}_{\pm\pm}$:}
  In the sector $\mc{H}_{\pm\pm}$ one has $p=c^2p_1+s^2p_2$
  and $\bar{p}=\bar{c}^2p_1-\bar{s}^2p_2$. Solving the
  equation $p=\bar{p}$ for $p_2$ and $p$ we obtain
\begin{equation}
  p_2\ =\ -\frac{c^2-\bar{c}^2}{s^2+\bar{s}^2}\,p_1\ =:\ tp_2
  \quad\text{ and }\quad
  p\ =\ \frac{c^2\bar{s}^2+s^2\bar{c}^2}{s^2+\bar{s}^2}\,p_1\ \ .
\end{equation}
  This is indeed consistent with the requirement $p_2>0$ as long as
  $\alpha>\balpha$. Similarly we have to solve the equation
\begin{equation}
  j\ :=\ j_1+j_2\pm\mu p_1\pm(\lambda-\mu)p_2\pm n
   \ = \ j_1-j_2\pm\mubb p_1\mp(\lambda-\mubb)p_2\mp\bar{n}
\end{equation}
  for $j_2$ and $j$. This yields
\begin{equation}
  \begin{split}
    j_2&\ =\ \mp(n+\bar{n})/2\mp\bigl[(\mu-\mubb)-(\mu+\mubb)t+2\lambda
    t\bigr]\,p_1/2\\[2mm]
    j&\ =\ j_1\pm(n-\bar{n})/2\pm\bigl[(\mu+\mubb)-(\mu-\mubb)t\bigr]\,p_1/2
  \end{split}
\end{equation}
  and completes the specification of the associated coset sector.
\bigskip

  \noindent{\em The sector $\mc{H}_{\pm\mp}$:}
  In the next sector one easily finds
  $p=c^2p_1-s^2p_2$ and
  $\bar{p}=\bar{c}^2p_1+\bar{s}^2p_2$. The usual procedure of
  equating $p$ and $\bar{p}$ results in
\begin{equation}
  p_2\ =\ -tp_1
  \quad\text{ and }\quad
  p\ =\ \frac{c^2\bar{s}^2+s^2\bar{c}^2}{s^2+\bar{s}^2}\,p_1\ \ .
\end{equation}
  This time we see the consistency with the assumption $\alpha<\balpha$.
  In addition to the previous equation we have to impose
\begin{equation}
  j\ :=\ j_1+j_2\pm\mu p_1\mp(\lambda-\mu)p_2\mp n
   \ = \ j_1-j_2\pm\mubb p_1\pm(\lambda-\mubb)p_2\pm\bar{n}\ \ .
\end{equation}
  Hence we immediately conclude that
\begin{equation}
  \begin{split}
    j_2&\ =\ \pm(n+\bar{n})/2\mp\bigl[(\mu-\mubb)-(\mu+\mubb)t+2\lambda t\bigr]\,p_1/2\\[2mm]
    j&\ =\ j_1\mp(n-\bar{n})/2\pm\bigl[(\mu+\mubb)-(\mu-\mubb)t\bigr]\,p_1/2\ \ .
  \end{split}
\end{equation}
\bigskip

  \noindent{\em The sector $\mc{H}_{\pm0}$:}
  In the decomposition of the sectors $\mc{H}_{\pm0}$ one obviously has
  $p=c^2p_1$ and $\bar{p}=\bar{c}^2p_1$ and therefore these sectors
  can only arise when
  $\alpha=\balpha$. The second constraint is
\begin{equation}
  j\ :=\ j_1+j_2\pm\mu p_1+n
   \ = \ j_1-j_2\pm\mubb p_1+\bar{n}\ \ .
\end{equation}
  Solving this for $j_2$ and plugging it back again results in
\begin{equation}
  j_2\ =\ -(n-\bar{n})/2\mp(\mu-\mubb)\,p_1/2
  \quad\text{ and }\quad
  j\ =\ j_1+(n+\bar{n})/2\pm(\mu+\mubb)\,p_1/2\ \ .
\end{equation}
We also have to require that $j_2$ lies in the interval $[0,1)$.
This restricts the parameters $n$ and $\bar n$. For instance when
$\m = \bar \m$ this implies $n = \bar n$ and $j_2 = 0$ or $n = \bar
n + 1$ and $j_2 = 1/2$.
\bigskip

  \noindent{\em The sector $\mc{H}_{00}$:}
  The discussion of the sector $\mc{H}_{00}$ parallels the one in the
  previous subsection. The only difference is in the constraint
\begin{equation}
  j\ :=\ j_1+j_2\  = \ j_1-j_2\mod\Integer \ ,
\end{equation}
  which has the solution
\begin{equation}
  j_2\ =\ \Upsilon\in\{0,1/2\}
  \quad\text{ and }\quad
  j\ =\ j_1+\Upsilon\mod\Integer\ \ .
\end{equation}
  In addition $s_{\text{min}} \le s \le s_{\text{max}}$ where
  $s_{\text{min}}$ and $s_{\text{max}}$ are given in Eq.~\eqref{eq:Borders}.

\subsection{The cosets of type $(++)$}

Our next goal is to determine the spectrum of the last class of models, the
$(++)$ cosets. When we considered their Lagrangian description in
section \ref{sc:Data} and \ref{sc:CosetGeometries} we found that the
$\s$-model fields $u_1$ and $u_2$ satisfy the constraint
\be
  U(z,\bar z)\ : =\ \bigl(\nu_1^2+\nubb_1^2\bigr) \ u_1(z,\bar z)
  + \bigl(\nu_2^2+\nubb_2^2\bigr) \ u_2(z,\bar z)\ = \rho \ \ ,
\ee
with $\rho \in \mathbb{R}$. We need to find a way to impose this
constraint on the spectrum of the original $H_4 \times H_4$ WZW
model. It is convenient to decompose the scalar field $U(z, \bar z)$
in its zero-mode, holomorphic and anti-holomorphic components
\be
  U(z, \bar z) = U_0 + U(z) + \bar U(\bar z) \ ,
\ee
so that the
previous constraint can be expressed in the form
\be
  U_0 = \rho \ ,
\hspace{1cm} U(z) = 0 \ , \hspace{1cm} \bar U(\bar z) = 0\ . 
\ee
Only the condition on the zero-mode $U_0$ correlates the
left and right Hilbert spaces of the original WZW model, while the
other two conditions can be imposed independently in the two Hilbert
spaces. In fact the derivatives of $U$ coincide with a linear
combination of the affine currents. More precisely $\p U(z) = -
\e\bigl(K(z)\bigr)$ and $\bar \p \bar U(\bar z) = \bar \e\bigl(\bar K(\bar z)\bigr)$
where $K(z)$ and $\bar K(\bar z)$ are the affine currents of the
$H_4$ subalgebra and $\e\bigl(K(z)\bigr) = \n_1^2 K^{(1)}(z) +  \n_2^2
K^{(2)}(z)$. For simplicity in this section we consider only the
case $\m = \bar \m = \lambda = 0$, which together with $\alpha =\bar
\alpha$ implies $\e = \bar \e$ .

The most efficient way to impose the constraint $\e\bigl(K(z)\bigr) = 0$ in
the holomorphic sector is to introduce ghost fields $(b,c)$ with
stress energy tensor $T_{gh} = - b \p c$ and conformal dimensions
$h_b = 1$ and $h_c = 0$. We then identify the physical Hilbert space
with the cohomology of the BRST charge \be Q = \oint \frac{dz}{2 \pi
i} c(z) \e\bigl(K(z)\bigr) \ . \ee As a result, the physical states are the
states $|\psi \rangle$ in the Hilbert space of the original WZW
model that satisfy the conditions
\be
  \e\bigl(K_{-n}\bigr) |\psi\rangle = 0 \
, \ \ \ \ n \ge 0 \qquad\text{ and }\qquad \e\bigl(J_{-n}\bigr)|\psi \rangle = 0 \ , \ \ \
\ n \ge 1 \label{eq:c++1} \ , \ee
where $\e\bigl(J(z)\bigr) = J^{(1)}(z) +
J^{(2)}(z)$. We proceed in exactly the same way in the
antiholomorphic sector, introducing ghost fields $(\bar b, \bar c)$
and a BRST charge $\bar Q$. This implies for the right modes of the
currents precisely the same conditions satisfied by the left modes.
Finally the constraint $U_0 = \rho$ leads to an additional condition
for the physical states
\be
  \left(\e\bigl(J_0\bigr) - \bar \e\bigl( \bar J_0\bigr) \right)|\psi \rangle\ =\ 0 \ \
  ,
  \label{eq:c++2}
\ee
 since $U_0$ and $\e\bigl(J_0\bigr) - \bar \e\bigl( \bar J_0\bigr)$ form a pair of
 canonical variables.
The constraints~\eqref{eq:c++1} can be solved only in the
$\mc{H}_{+-}$, $\mc{H}_{-+}$ and $\mc{H}_{00}$ sectors of the
original WZW model and therefore the spectrum is given by
\ba &&        \bigl[(\pm|p_1,j_1),(\mp|p_2,j_2);(0|s,j)\bigr] \nb \\
&& \bigl[(0|s_1,j_1),(0|s_2,j_2);(0|s,j)\bigr]  \label{eq:schema++}
\ea
  with $p_2 = \cot^2 \alpha \, p_1$. Note that if we had followed
exactly the same approach as in the previous subsections and imposed
the constraint \eqref{eq:cosetconstraint} with $X = K_0$, we would
have reached the conclusion that every sector $\mc{H}_{\m \n}$
contributes to the spectrum with no restrictions on the labels $p_1$
and $p_2$.

We still have to require the invariance of the physical spectrum
with respect to the residual gauge transformations generated by the
modes of the affine currents $P_i(z)$, $i = 1, 2$, and by the zero
mode of the current $\e\bigl(J(z)\bigr) + \bar \e\bigl(\bar J(\bar z)\bigr)$. This is
equivalent to the requirement that the constraints in
\eqref{eq:cosetconstraint} are satisfied for this restricted set of
generators.

Before imposing these conditions, we would like to make a few
comments about the energy-momentum tensor of the coset model. Let us
start with the energy-momentum tensor associated with the product of the $H_4 \times H_4$ model
and the ghost fields
\be
  T_{\text{tot}}(z)\ =\ T_{H_4}^{(1)}(z)+T_{H_4}^{(2)}(z) + T_{gh}(z) \ .
\ee
  This energy-momentum tensor has central charge $c = 6$ and when restricted to the cohomology it can be written as \be
T_{\text{tot}}(z) \sim T_{H_4}^{(1)}(z)+T_{H_4}^{(2)}(z) - \e\bigl(J(z)\bigr) 
\e\bigl(K(z)\bigr) \label{emem}
\ee
  due to the relation
\be
\left \{Q, \e\bigl(J(z)\bigr) b(z) \right \} = T_{gh}(z) + \e\bigl(J(z)\bigr) \e\bigl(K(z)\bigr) \ . \ee
  The central charge is further reduced
to $c=4$ by the gauging of the affine currents $P_i(z)$, $i = 1, 2$.
This can be accomplished by subtracting  from \eqref{emem} the $c = 2$
energy-momentum tensor \be T_r(z) = \frac{1}{2} \left [ \e\bigl(P_1(z)\bigr)^2
+ \e\bigl(P_2(z)\bigr)^2 + \e\bigl(K(z)\bigr)^2 \right ] \ .
\ee
As a result in
the physical subspace the stress-energy tensor of this class of
models coincides with the one given by the standard coset
construction \cite{Bardakci:1971nb, Goddard:1985vk} \be T_{(++)} =
T_{H_4}^{(1)}+T_{H_4}^{(2)} - T_{\e\bigl(H_4\bigr)}   \ .
\ee

  It remains to determine the relations among the labels of the
representations in \eqref{eq:schema++}.
\bigskip

{\noindent}{\em The sector $\mc{H}_{\pm\mp}$:} This sector
decomposes into continuous representations. Here we find the
constraint \be j\ = \ j_1+j_2  \ \ \ \ \mod\Integer \  ,
\hspace{1cm} \bar \jmath = \ j_1+j_2 \ \ \ \ \mod\Integer \ ,
\label{jpp} \ee with $j = 0$ and $\bar \jmath = 0$. The equation can
be solved by \be j_2\
 = \ -j_1 + n/2  \ , \hspace{1cm}
j = n/2  \ \ \ \ \ \mod\Integer \ ,
\ee with $n \in \mathbb{Z}$.
\bigskip

  \noindent{\em The sector $\mc{H}_{00}$:}
  In this sector we have  the
  constraint $j_1 + j_2 = 0 \ \mod\Integer$ with solution
  $j_2=1-j_1$ and $j= 0$. The range of $s$ is
  $|\n_1 s_1 - \n_2 s_2|  \, \leq s \leq \,  \n_1 s_1 +  \n_2 s_2$.

\section{\label{sc:conclusions}Conclusions}

In this article we studied in detail the diagonal cosets of the
Heisenberg group. These cosets form a large and interesting class of curved
string theory backgrounds and provide an example of a coset
construction where both the numerator and the denominator group are
non-compact and non-abelian. We classified all possible diagonal
cosets and derived the metric, dilaton and antisymmetric tensor that
specify the corresponding curved space-times. We found three classes
of models, thereby generalising the results of
\cite{Antoniadis:1994jx}. The resulting models are all particular
examples of a family of string backgrounds  related to plane waves
by abelian T-duality transformations \cite{Klimcik:1993kd}.

Our three classes of models are labelled by two signs and are called
$(++)$, $(+-)$ and $(--)$, respectively. A minus sign means that the
right embedding includes a certain twist automorphism
acting on the corresponding $H_4$ factor in $H_4 \times H_4$. All
three classes depend on several continuous parameters, in particular
on two angles $\alpha$, $\bar \alpha$. The models considered in
\cite{Antoniadis:1994jx} correspond to the symmetric case $\alpha =
\bar \alpha$ and describe singular space-times. The general models
with $\alpha \ne \bar \alpha$ introduced in this paper are
generically non-singular and interpolate between singular and
non-singular space-times.

  Apart from the geometric description we also derived the spectrum of
  the diagonal cosets using conformal field theory techniques. In
  order to achieve this we first studied the decomposition of the
  tensor products of affine $H_4$ representations with respect to the
  embedded $H_4$ algebra. We described a method to derive the
  branching functions for representations with zero spectral flow
  which avoided the problems that would have arisen if we had tried to
  use the standard technique of character decompositions. We expect
  that the approach followed in this paper will be useful for the
  treatment of other non-compact and non-abelian cosets such as
  e.g.\ $SL(2,\Real)\times SL(2,\Real)/SL(2,\Real)$.

  Several aspects of the models discussed in this paper deserve
  further investigation. First of all we would like to find a rigorous
  method to study the decomposition of the tensor products of spectral
  flow representations, for which we could only present partial
  results. In this way one would obtain a complete description of the
  spectrum of the diagonal cosets and could also study their one-loop
  partition functions.

  Following the work of Antoniadis and Obers \cite{Antoniadis:1994jx}
  (see also \cite{Kiritsis:1994hg}) we would also like to investigate
  the effect of T-duality transformations on the space of the diagonal
  cosets of the Heisenberg group and in particular to consider their
  action on the spectrum of the models described in this paper. In
  this way one could establish under which conditions the duality
  symmetries of the curved backgrounds reflect exact symmetries of the
  underlying coset conformal field theories.

  The particular example considered in this paper shows that when the
  coset construction is applied to a non-semisimple group there is a
  significant amount of freedom, to the point that the resulting coset
  conformal field theories usually come in continuous families. This
  should be a  generic feature of non-semisimple cosets and it would
  be worth exploiting it to construct other models of this type. For
  instance it would be very interesting to find other four-dimensional
  models that could be identified with families of curved string
  backgrounds, as it is the case for the diagonal cosets of the
  Heisenberg group. A possible class of this type are the cosets
  $(H_4)^{n+1}/(H_4)^n$.

  Finally, another valuable line of research would be the study of
  string interactions in these backgrounds. With the information
  gathered in this paper it should be possible to construct
  correlation functions of coset primary fields. In order to determine
  the three- and four-point couplings one should use the structure
  constants of the $H_4$ WZW model derived in
  \cite{D'Appollonio:2003dr} and properly generalise the analysis
  performed recently in~\cite{D'Appollonio:2007bn} for the abelian
  cosets.

\acknowledgments
  The authors are grateful to Jan de\,Boer, Terry Gannon, Wolter
  Groenevelt, Sylvain Ribault, Volker Schomerus, Jasper Stokman and
  J\"org Teschner fur useful discussions. T.Q.\ would like to thank
  King's College London and the Isaac Newton Institute for
  Mathematical Sciences in Cambridge for their kind hospitality during
  the last stages of preparing this article. G.D.\ acknowledges the
  support of the PPARC rolling grant PP/C507145/1. The research of
  T.Q.\ is funded by a Marie Curie Intra-European Fellowship, contract
  number MEIF-CT-2007-041765. We furthermore acknowledge partial
  support from the EU Research Training Network {\it Superstring
    theory}, MRTN-CT-2004-512194.

\newpage

\appendix
\section{\label{ap:FullGeometries}Geometries for the fully asymmetric cosets of type $(--)$ and $(+-)$}

  For the sake of completeness we reserved this appendix to summarise
  the geometric data that arise for the fully asymmetric versions of
  the cosets of type $(--)$ and $(+-)$.

\subsection{Cosets of type $(--)$}

  Before we spell out the background data for the fully
  asymmetric coset of type $(--)$ it is convenient to define the
  quantity
\begin{equation}
  \begin{split}
    r(u)\ =\ 2c^2\bar{c}^2-c^2-\bar{c}^2+2c\bar{c}s\bar{s}\cos(2u)\ \ ,
  \end{split}
\end{equation}
  where we used the notation introduced in Eq.~\eqref{eq:shorts}.
  The function $r(u)$ is a
  generalisation of the functions $r^\pm(u)$ and $R^\pm(u)$ defined in
  the main text. It completely specifies the dilaton
\begin{equation}
  \Phi\ =\ -\frac{1}{2}\ln\bigl(r(u)\bigr)\ \ .
\end{equation}
  Employing the previous definition one can also simplify the metric
  which assumes the form
\begin{equation}
  \begin{split}
    ds^2
    &\ =\ 4dudv
          +\frac{1}{r(u)}\Bigl\{
           A_{xx}dx^2-2A_{xy}dxdy+A_{yy}dy^2
          +2\bigl(B_xx+B_yy\bigr)\sin(u)dxdu\\[2mm]
    &\qquad+2\bigl(C_xx+C_yy\bigr)\sin(u)dydu
          +\bigl(D_{xx}x^2+D_{xy}xy+D_{yy}y^2\bigr)du^2
          \Bigr\}+D\,du^2\ \ .
  \end{split}
\end{equation}
  In order to keep this expression short we list the auxiliary
  functions appearing in this expression separately. The first group
  of functions is given by
\begin{equation}
  \begin{split}
    A_{xx}
    &\ =\ \Bigl\{c^2+\bar{c}^2-2c^2\bar{c}^2+2c\bar{c}s\bar{s}-2
             -2\bigl(c\bar{c}+s\bar{s}+2c\bar{c}s\bar{s}\cos(u)\bigr)\cos(u)\Bigr\}\\[2mm]
    A_{xy}
    &\ =\ \Bigl\{c^2+3\bar{c}^2-2+2\left(c\bar{c}^3-s\bar{s}^3\right)\cos(u)\Bigr\}\\[2mm]
    A_{yy}
    &\ =\ -\Bigl\{2+4\bar{c}^4-5 \bar{c}^2
           +2c\bar{c}s\bar{s}
           +c^2\bigl(2\bar{c}^2-1\bigr)\\[2mm]
    &\hspace{3cm}+2\left(2\bar{c}^2-1\right)\left(c\bar{c}-s\bar{s}\right)\cos(u)
           -4c\bar{c}s\bar{s}\cos^2(u)\Bigr\}\ \ .
  \end{split}
\end{equation}
  The second and the third group have the following form,
\begin{equation}
  \begin{split}
    B_x
    &\ =\ \bigl(c^2+\bar{c}^2-1\bigr) \bigl(c \bar{c}-s
    \bar{s}\bigr)\\[2mm]
    B_y
    &\ =\ \Bigl\{c\bar{c}\left(-2\bar{c}^4+2\bar{c}^2+c^2\right)
           +s\bar{s}\left(2\bar{c}^4-2\bar{c}^2+c^2-1\right)
           +2c\bar{c}s\bar{s}\left(2 c^2-1\right)\cos(u)\Bigr\}\\[2mm]
    C_x
    &\ =\ \Bigl\{
            \Bigl(1-2 \bar{c}^4+\bar{c}^2+c^2\bigl(2\bar{c}^2-1\bigr)\Bigr)
            \bigl(c\bar{c}+s\bar{s}\bigr)
            +4c\bar{c}s\bar{s}\left(2\bar{c}^4+\left(2 c^2-3\right)\bar{c}^2-c^2\right)\cos(u)\Bigr\}\\[2mm]
    C_y
    &\ =\ \bigl(2\bar{c}^2-1\bigr)\Bigl\{c^2(\bar{c}c-s\bar{s})
           +s\bar{s}-2c\bar{c}s\bar{s}\cos(u)\Bigr\}\ \ .\raisetag{20pt}
  \end{split}
\end{equation}
  Finally we have
\begin{equation}
  \begin{split}
    D_{xx}
    &\ =\ -\frac{1}{4}
           \left(2c^2-1\right)
           \left(2\bar{c}^2-1\right)
           \Bigl\{2+2c^2\bar{c}^2-c^2-\bar{c}^2
                 -2c\bar{c}s\bar{s}\\[2mm]
    &\hspace{5.5cm}-2\left(c\bar{c}+s\bar{s}\right)\cos(u)
                 +4c\bar{c}s\bar{s}\cos^2(u)
           \Bigr\}\\[2mm]
    D_{xy}
    &\ =\ \frac{1-2c^2}{2}\Bigl\{
           2-2 \bar{c}^4+\bar{c}^2+c^2\left(2\bar{c}^2-1\right)
           -2\Bigl(c\bar{c}^3\bigl(3-2\bar{c}^2\bigr)
           +s\bar{s}^3\bigl(3-2\bar{s}^2\bigr)\Bigr)\cos(u)\Bigr\}\\[2mm]
    D_{yy}
    &\ =\ -\frac{1-2c^2}{4}\Bigl\{
           2+8\bar{c}^6+12c^2\bar{c}^4-18\bar{c}^4-12c^2\bar{c}^2
           -c^2+7\bar{c}^2
           +2c\bar{c}s\bar{s}\bigl(1-2\bar{c}^2\bigr)\\[2mm]
    &\hspace{3.5cm}+2\bigl(c\bar{c}-s\bar{s}\bigr)
           \bigl(4\bar{c}^2\bar{s}^2+1\bigr)\cos(u)
           -4c\bar{c}s\bar{s}\bigl(1-2\bar{c}^2\bigr)\cos^2(u)\Bigr\}\\[2mm]
    D
    &\ =\ 2\Lambda\Bigl\{
          \lambda\bigl(4c^2\bar{c}^2-c^2-\bar{c}^2\bigr)
          +\mu\bigl(1-2\bar{c}^2\bigr)
          +\bar{\mu}\bigl(1-2c^2\bigr)\Bigr\}\ \ .\raisetag{20pt}
  \end{split}
\end{equation}
  For a general choice of parameters the background also support a
  three-form flux. Using the same conventions as above it may be
  expressed as
\begin{equation}
  \begin{split}
    H&\ =\ \frac{2\bar{c}\bar{s}}{r(u)^2}\bigl(c^2-\bar{c}^2\bigr)\Bigl\{
           \bar{c}s\bigl(4\bar{c}^2c^2-c^2-\bar{c}^2+2s^2\bigr)
           +c\bar{s}\bigl(4\bar{c}^2c^2-c^2-\bar{c}^2+2\bar{s}^2\bigr)\\[2mm]
     &\qquad+2cs\bigl(2\bar{c}^2c^2+2-c^2-\bar{c}^2-2c\bar{c}s\bar{s}\bigr)\cos(u)
           +4c\bar{c}s\bar{s}\left(\bar{c}s+c\bar{s}\right)\cos^2(u)\\[2mm]
     &\qquad+8c^2\bar{c}s^2\bar{s}\cos^3(u)\Bigr\}\sin(u)\,dx\wedge dy\wedge du\ \ .
  \end{split}
\end{equation}
  The background described here may be cast into the standard
  form~\eqref{eq:StdPW} of a gravitational plane wave by a suitable
  change of coordinates. Since  one has to follow a rather cumbersome
  procedure in order to find the explicit coordinate transformation,
  we refrain from doing so.

\subsection{Cosets of type $(+-)$}

  Let us turn our attention to the third type of cosets of class
  $(+-)$ now. This time a crucial ingredient of the metric and the
  other background fields are the following functions
\begin{equation}
  r_c^\pm(u)\ =\ c^2+\bar{c}^2\pm2c\bar{c}\cos(u)\ \ ,\qquad
  r_s^\pm(u)\ =\ s^2+\bar{s}^2\pm2s\bar{s}\cos(u)\ \ .
\end{equation}
  Again we used the same abbreviations as in \eqref{eq:shorts}. Like before
  the auxiliary functions are useful in order to express the dilaton which is
  given by
\begin{equation}
  \Phi\ =\ -\frac{1}{2}\ln\bigl(r_c^-(u)\bigr)
\end{equation}
  The metric is also easily derived. Its shape resembles the one for
  the $(--)$-gauging and its explicit form is given by
\begin{equation}
  \label{eq:PMAsymmetric}
  \begin{split}
    ds^2
    &\ =\ \frac{2dudv}{s^2}
          +\frac{1}{r_c^-(u)}\Bigl\{
           A_{xx}dx^2-2A_{xy}dxdy+A_{yy}dy^2
           +\bigl(B_xx+B_yy\bigr)\sin(u)dxdu\\[2mm]
    &\qquad+\bigl(C_xx+C_yy\bigr)\sin(u)dydu
           +\bigl(D_{xx}x^2+D_{xy}xy+D_{yy}y^2\bigr)du^2
          \Bigr\}
          +D\,du^2\ \ .
  \end{split}
\end{equation}
  The main difference to the $(--)$-case can be found in the auxiliary
  functions needed to express eq.\ \eqref{eq:PMAsymmetric}. For the
  first set of functions one obtains
\begin{equation}
  \begin{split}
    A_{xx}
    &\ =\ r_s^+(u)\\[2mm]
    A_{xy}
    &\ =\ \Bigl\{
           2+\bigl(2\bar{c}^2-1\bigr)c^2
           +2c\bar{c}s\bar{s}-3\bar{c}^2
           -2\bar{s}^2\bigl(c\bar{c}-s\bar{s}\bigr)\cos(u)
           -4c\bar{c}s\bar{s}\cos^2(u)\Bigr\}\\[2mm]
    A_{yy}
    &\ =\ -\Bigl\{
          -2+8c^2\bar{c}^4-4\bar{c}^4-8 c^2 \bar{c}^2+5 \bar{c}^2
          +4c\bar{c}s\bar{s}\bigl(2\bar{c}^2-1\bigr)+c^2\\[2mm]
    &\hspace{2.5cm}-2\Bigl(s\bar{s}\bigl(1-2\bar{c}^2\bigr)-2c\bar{c}\bar{s}^2\Bigr)\cos(u)
          -8c\bar{c}s\bar{s}\bigl(2\bar{c}^2-1\bigr)\cos^2(u)\Bigr\}\ \ .
  \end{split}
\end{equation}
  The second set is given by
\begin{equation}
  \begin{split}
    B_x
    &\ =\ \frac{1}{s^2}\Bigl\{
          \bigl(1-2\bar{c}^2\bigr)c^3\bar{c}
          +\bigl(3-2\bar{c}^2\bigr)s\bar{s}c^2
          +\bar{c}^3 c+s\bar{s}\bigl(\bar{c}^2-2\bigr)\Bigr\}\\[2mm]
    B_y
    &\ =\ \frac{1}{s^2\bar{s}^2}\Bigl\{
          \Bigl(-4 \bar{c}^4+7 \bar{c}^2+c^2 \left(4 \bar{c}^4-8
   \bar{c}^2+3\right)-2\Bigr)
          \left(c \bar{c}-s \bar{s}\right)
          +4c\bar{c}s\bar{s}^3\bigl(2 c^2-1\bigr)\cos(u)\Bigr\}\\[2mm]
    C_x
    &\ =\ -\frac{1}{s^2\bar{s}^2}\Bigl\{
           \Bigl(-2 \bar{c}^4+3 \bar{c}^2+c^2 \left(4 \bar{c}^4-6
   \bar{c}^2+3\right)-2\Bigr)
           \bigl(c\bar{c}+s\bar{s}\bigr)
           +4c\bar{c}s^3\bar{s}\cos(u)\Bigr\}\\[2mm]
    C_y
    &\ =\ -\frac{1}{s^2\bar{s}^2}\Bigl\{
           c\bar{c}\bigl(2\bar{c}^4-3\bar{c}^2+c^2\bigr)
           +s\bar{s}\Bigl(2+6\bar{c}^4-9\bar{c}^2+c^2\bigl(-8\bar{s}^4+4\bar{s}^2+1\bigr)\Bigr)\\[2mm]
    &\hspace{3cm}+4c\bar{c}s\bar{s}\Bigl(-6 \bar{c}^4+9 \bar{c}^2+c^2
    \left(8 \bar{c}^4-12 \bar{c}^2+3\right)-2\Bigr)\cos (u)\Bigr\}\ \ .\raisetag{20pt}
  \end{split}
\end{equation}
  Finally, the remaining ones assume the form
\begin{equation}
  \begin{split}
    D_{xx}
    &\ =\ \frac{\bigl(1-2c^2\bigr)\,r_s^-(u)}{4s^2\bar{s}^2}\\[2mm]
    D_{xy}
    &\ =\ -\frac{1-2c^2}{2s^2}\Bigl\{
           2+4 \bar{c}^4-7 \bar{c}^2-2c\bar{c}s\bar{s}+c^2\bigl(2\bar{c}^2-1\bigr)\\[2mm]
    &\hspace{3cm}-2\left(2\bar{c}^4-3\bar{c}^2+1\right)\bigl(c\bar{c}+s\bar{s}\bigr)\cos(u)
           +4c\bar{c}s\bar{s}\cos^2(u)\Bigr\}\\[2mm]
    D_{yy}
    &\ =\ \frac{1-2c^2}{4s^2}\Bigl\{
           2-c^2-13\bar{c}^2+16c^2\bar{c}^6-8 \bar{c}^6-32 c^2 \bar{c}^4+20\bar{c}^4+16 c^2 \bar{c}^2\\[2mm]
    &\hspace{1cm}-4c\bar{c}s\bar{s}\left(4\bar{c}^4-6\bar{c}^2+1\right)
           -2\Bigl(2c\bar{c}\bigl(2\bar{c}^4-3\bar{c}^2+1\bigr)
             +s\bar{s}\bigl(4\bar{c}^4-6
             \bar{c}^2+1\bigr)\Bigr)\cos(u)\\[2mm]
    &\hspace{5cm}+8c\bar{c}s\bar{s}\left(4\bar{c}^4-6\bar{c}^2+1\right)\cos^2(u)\Bigr\}\\[2mm]
    D
    &\ =\ \frac{\Lambda}{s^2\bar{s}^2}\Bigl\{
           \lambda\bigl(2s^2\bar{c}^2-c^2-\bar{c}^2\bigr)
           +\mu-\mubb\bigl(1-2c^2\bigr)\Bigr\}\ \ .\raisetag{20pt}
  \end{split}
\end{equation}
  The gauged WZW model also comes with a non-trivial three-form
  flux which is needed in order to ensure conformal invariance. A
  straightforward calculation yields
\begin{equation}
  H\ =\ \frac{dx\wedge dy\wedge du}{s^2\bar{s}^2\,r_c^-(u)^2}
        \bigl(c^2-\bar{c}^2\bigr)
        \Bigl(2+2c^2\bar{c}^2-r_c^+(u)\Bigr)
        \Bigl(c\bar{c}\bar{s}^2-\bar{c}^2s\bar{s}+2c\bar{c}s\bar{s}\cos(u)\Bigr)\sin(u)\ \ .
\end{equation}

\section{\label{ap:AdTensor}Tensor products of $H_4$-representations}

  In this appendix we summarise a few tensor products of finite and
  infinite dimensional representations of the Heisenberg Lie algebra
  $H_4$. We shall show that whenever the generator $K$ vanishes,
  reducible but indecomposable representations can appear in the
  tensor product.

\subsection{Adjoint times adjoint}

  The adjoint representation of $H_4$ mirrors the non-semi-simplicity
  of the Lie algebra. Its structure may be read off from the
  composition series
\begin{equation}
  \text{ad}:\quad
  \xymatrixrowsep{8pt}\xymatrixcolsep{8pt}
  \begin{array}{c}
  \xymatrix{
    & [1] \ar[dr]& &\\
    [0] \ar[dr] \ar[ur] & & [0]\ \ .\\
    & [-1] \ar[ur]
  }
  \end{array}
\end{equation}
  In this diagram $[j]$ denotes a one-dimensional representation on
  which $-iJ$ acts as the scalar $j$ while $K$ acts trivially. To the
  right we find an irreducible invariant subspace $[0]$, given by the
  span of $K$. If we divide out this subspace we find two new
  invariant subspaces, $[1]$ and $[-1]$, represented by
  $P^\pm$. Taking again the corresponding quotient we finally end up
  with a second $[0]$, the span of $J$. Similar diagrams will be used
  for the more complicated representations discussed below.
\smallskip

  We are interested in (symmetrised) tensor products of the adjoint
  representation with itself since they are relevant for the
  construction of the affine modules. The main contribution to
  $\ad\otimes\ad$ is schematically given by
\begin{equation}
  \xymatrixrowsep{8pt}\xymatrixcolsep{8pt}
  \xymatrix{
    && P^+\otimes P^+ \ar[dr] \\
    & \hspace{-0.5cm}J\otimes P^++P^+\otimes J \ar[ur] \ar[dr] &&
    \hspace{-0.5cm}K\otimes P^++P^+\otimes K \ar[dr]\\
    J\otimes J \ar[dr] \ar[ur]&&\dots \ar[ur] \ar[dr] && K\otimes K \ , \\
    & \hspace{-0.5cm}J\otimes P^-+P^-\otimes J \ar[dr] \ar[ur] &&
    \hspace{-0.5cm}K\otimes P^-+P^-\otimes K \ar[ur]\\
    && P^-\otimes P^- \ar[ur]
   }
\end{equation}
  which is a nine-dimensional indecomposable representation.
  It is part of an infinite series of indecomposable
  representations which arise in higher tensor products of the adjoint
  representation with itself. In fact if we start with the
  state $J\otimes J\otimes\cdots\otimes J$ on the left hand side, we expect
  to find a representation of dimension $n^2$ whose
  schematic description is
\begin{equation}
  \xymatrixrowsep{8pt}\xymatrixcolsep{8pt}
  \begin{array}{c}
  \xymatrix{
    && [n] \ar[dr] \\
    & \cdots \ar[ur] \ar[dr] && \cdots \ar[dr]\\
    [0] \ar[dr] \ar[ur]&&\cdots \ar[dr] \ar[ur] && [0]\\
    & \cdots \ar[dr] \ar[ur] && \cdots \ar[ur]\\
    && [-n] \ar[ur]
  }
  \end{array}
  \ \cong\ %
  \begin{array}{c}
  \xymatrix{
    && (P^+)^{\otimes n} \ar[dr] \\
    & \cdots \ar[ur] \ar[dr] && \cdots \ar[dr]\\
    J^{\otimes n} \ar[dr] \ar[ur]&&\cdots \ar[dr] \ar[ur] && K^{\otimes n}\\
    & \cdots \ar[dr] \ar[ur] && \cdots \ar[ur]\\
    && (P^-)^{\otimes n} \ar[ur]
  }
  \end{array} \ .
\end{equation}
  The product $\ad\otimes\ad$ also contains a singlet given by
\begin{equation}
  [0]\ \cong\ %
  2(K\otimes J+J\otimes K)+(P^+\otimes P^-+P^-\otimes P^+)\ \ ,
\end{equation}
 which cannot be
  reached from any other state. The remaining six vectors belong to the antisymmetric part
  of the tensor product $\ad\otimes\ad$. They form two
  three-dimensional indecomposable representations which have the
  structure
\begin{equation}
  \xymatrixrowsep{0pt}\xymatrixcolsep{16pt}
  \begin{array}{c}
  \xymatrix{
    && [1] \\
    & [0] \ar[ur] \\
    [-1] \ar[ur] &&
  }
  \end{array}
  \qquad\text{ and }\qquad
  \begin{array}{c}
  \xymatrix{
    [1] \ar[dr] &&\\
    & [0] \ar[dr] &\\
    && [-1]
  }
  \end{array}\ \ .
\end{equation}

\subsection{Continuous times adjoint}

  We would like to show that the tensor product $(0|s,j)\otimes\ad$ is
  indecomposable. This observation is not particularly surprising
  since the generator $K$ acts trivially in both constituents and
  consequently on the whole module. Nevertheless the statement is
  non-trivial and has to be checked thoroughly. We will argue that the
  Casimir operator is not diagonalisable on the tensor product, thus
  proving our assertion.
\smallskip

  Let us consider the four-dimensional subspace of vectors with $-iJ=j+n$
  ($n\in\Integer$). Denote by $|v\rangle$ the vector in $(0|s,j)$ with
  $-iJ|v\rangle=(j+n)|v\rangle$. A convenient basis is then given by
  the linear combinations
\begin{align}
  v_1&\ =\ P^+|v\rangle\otimes P^-+P^-|v\rangle\otimes P^+\ \ ,&
  v_2&\ =\ -4s^2\,|v\rangle\otimes K\\[2mm]
  v_3&\ =\ -|v\rangle\otimes K-i\bigl(P^+|v\rangle\otimes
           P^--P^-|v\rangle\otimes P^+\bigr)\ \ ,&
  v_4&\ =\ |v\rangle\otimes J\ \ .
\end{align}
  The total quadratic Casimir may be expressed as
\begin{equation}
  \begin{split}
    C&\ =\ \frac{1}{2}(P^+P^-+P^-P^+)+2JK\\[2mm]
     &\ =\ \frac{1}{2}\bigl[(P_1^++P_2^+)(P_1^-+P_2^-)+(P_1^-+P_2^-)(P_1^++P_2^+)\bigr]
     +2(J_1+J_2)(K_1+K_2)\\[2mm]
     &\ =\ C_1+C_2+P_1^+P_2^-+P_1^-P_2^++2J_1K_2+2K_1J_2
  \end{split}
\end{equation}
  in terms of the Casimirs and of the generators of the individual algebras.
  For the vectors above we have $C_1=s^2$ and $K_1=K_2=0$. This
  simplifies the calculations considerably and leads to the
  following matrix form,
\begin{equation}
  C\ =\ \mat s^2&0&0&0\\0&s^2&1&0\\0&0&s^2&1\\0&0&0&s^2\tam\ \ .
\end{equation}
  We thus proved that the affine continuous representation is not
  completely reducible with respect to its horizontal subalgebra.
  Similar indecomposable representations will appear on higher
  energy levels but we leave the complete analysis for future
  work. Note that the occurrence of indecomposable representations in
  the CFT should not affect the string theory spectrum, since the
  states created by the negative modes of the currents $J$ and $K$ are
  not physical, i.e.\ they are removed by the Virasoro constraints.

As we said indecomposable representations only appear when $K = 0$,
so we expect that the tensor product
  $(+|p,j)\otimes\ad$ (with $p\neq0$)
is completely reducible. It is easy to prove that in fact
\begin{equation}
  (+|p,j)\otimes\ad\ =\ (+|p,j+1)\oplus2(+|p,j)\oplus(+|p,j-1)\ \ .
\end{equation}
To see this let us assume
  that the infinite dimensional discrete representation is
  generated by a vector $|v\rangle$ with
\begin{align}
  P^+|v\rangle&\ =\ 0&
  K|v\rangle&\ =\ ip|v\rangle&
  J|v\rangle&\ =\ ij|v\rangle\ \ .
\end{align}
  Then it is not difficult to find four highest weight, none of which leads to non-trivial invariant
  subspaces.\footnote{Note that in this respect $H_4$ differs from the
    case of $AdS_3$ or $SL(2,\Real)$ where representations arise that
    are not fully reducible, e.g.\ in the tensor product $(+,1)\otimes
    1$.} The corresponding highest weight vectors read
\begin{align}
  |v\rangle\otimes P^+ \ ,&&
  |v\rangle\otimes K \ , &&
  P^-|v\rangle\otimes P^++2pi|v\rangle\otimes J \ , &&
  P^-|v\rangle\otimes K-ip|v\rangle\otimes P^-\ \ .
\end{align}
  For convenience we ordered the highest weight states by their
  eigenvalues with respect to $-iJ$. In the given order the latter
  read $j+1$, $j$, $j$ and $j-1$.


\def\cprime{$'$} \def\cprime{$'$}
\providecommand{\href}[2]{#2}\begingroup\raggedright\endgroup

\end{document}